\documentclass[11pt]{JHEP3}
\usepackage{epsfig}

\newcommand{\beq}{\begin{eqnarray}}
\newcommand{\eeq}{\end{eqnarray}}
\newcommand{\centeron}[2]{{\setbox0=\hbox{#1}\setbox1=\hbox{#2}\ifdim                                       
\wd1>\wd0\kern.5\wd1\kern-.5\wd0\fi \copy0
\kern-.5\wd0\kern-.5\wd1\copy1\ifdim\wd0>\wd1 \kern.5\wd0\kern-.5\wd1\fi}}
\newcommand{\ltap}{\>\centeron{\raise.35ex\hbox{$<$}}
                               {\lower.65ex\hbox{$\sim$}}\>}
\newcommand{\gtap}{\>\centeron{\raise.35ex\hbox{$>$}}
                               {\lower.65ex\hbox{$\sim$}}\>}

\newcommand\ZZ{\hbox{\zfont Z\kern-.4emZ}}
\font\zfont = cmss10 

\title{ Bounds and Decays of New Heavy Vector-like Top Partners} 

\author{Giacomo Cacciapaglia, Aldo Deandrea\\ Universit\'e de Lyon, France; Universit\'e Lyon 1,\\
   CNRS/IN2P3, UMR5822 IPNL, F-69622 Villeurbanne Cedex, France\\ E-mail: \email{ g.cacciapaglia@ipnl.in2p3.fr, deandrea@ipnl.in2p3.fr}}
\author{Daisuke Harada, Yasuhiro Okada\\ KEK Theory Center, Institute of Particle and Nuclear Studies, KEK, 
1-1 Oho, Tsukuba, Ibaraki 305-0801, Japan\\ Department of Particle and Nuclear Physics, Graduate
University for Advanced Studies (Sokendai), 1-1 Oho, Tsukuba, Ibaraki 305-0801, Japan\\ E-mail: \email{dharada@post.kek.jp, yasuhiro.okada@kek.jp}}
\abstract{We study the phenomenology of new heavy vector-like fermions that couple to the third generation quarks via Yukawa interactions, covering all the allowed representations under the standard model gauge groups.
 We first review tree and loop level bounds on these states. We then discuss tree level decays and loop-induced decays to photon or gluon plus top. The main decays at tree level are to $W b$ and/or $Z$ and Higgs plus top via the new Yukawa couplings. The radiative loop decays turn out to be quite close to the naive estimate: in all cases, in the allowed perturbative parameter space, the branching ratios are mildly sensitive on the new Yukawa couplings and small. We therefore conclude that the new states can be observed at the LHC and that the tree level decays can allow to distinguish the different representations. Moreover, the observation of the radiative decays at the LHC would suggest a large Yukawa coupling in the non-perturbative regime.}
\preprint{ KEK-TH-1371, LYCEN 2010-06}
\begin{document}
\section{Introduction}
\label{sec:intro}
\setcounter{equation}{0}
\setcounter{footnote}{0}

In many models of new physics, like for example extra dimensional models, Little Higgs 
models, dynamical models, there are heavy vector-like fermions which decay to Standard Model (SM) fermions plus a boson 
($W$/$Z$ and/or Higgs $h$). Moreover the mixing of vector-like quarks with the third generation and in particular the top quark 
is a common feature in little Higgs models \cite{lhmod} and composite Higgs models based on top condensation \cite{tcmod}. 
We are at present at the beginning of the Large Hadron Collider (LHC) era which is an exciting time
for discovery of new particles and test of models near the electroweak scale. Previous collider and precision data place however 
limits on new heavy quarks and set the lowest mass scale for these resonances once some properties for these particles are 
assumed. Direct searches give mass constraints in the range of 200-300 GeV, typically assuming a charged current decay chain 
\cite{pdg}.  Precision tests can be stringent in some cases but more model dependent as the effect of new particles or couplings 
may affect the precision observables in both directions. Indeed in the past vector fermion models were for example suggested to 
improve the $R_b$ data at LEP2 \cite{Ma:1995zh,Bhattacharyya:1995ce,Chang:1996pf}. More recently it was proposed that the discrepancy of the measured asymmetry $A_b^{FB}$ with the standard model can be reduced through the introduction of new quarks with non-trivial mixings with the third generation \cite{Choudhury:2001hs}.
Mixing effects with the SM quarks give stringent bounds in the case of mixing with the first two generations but only mild bounds 
for the mixing with the third generation. For a discussion of the parametrisation of mixing effects and CP bounds see \cite{Branco:1986my}. In the following we shall focus on vector-like quarks in various representations for which a coupling to a standard Higgs 
doublet is possible. We will assume that there is only one fermion that couples to the third generation of quarks, the top and 
bottom, via a Yukawa coupling.
This situation can reproduce with a good accuracy models where only the top partners are lighter than other heavy fermions like, 
for example, composite Higgs models\cite{modHF}. 
This case is in general different from the case of a chiral fourth generation, for which more stringent bounds can be obtained 
\cite{Kribs,MH}. 

In order to keep the discussion general we will not consider any specific model but rather an effective approach 
where the decays are induced by a new Yukawa coupling. This coupling generates the mixing of the new heavy fermion with top and bottom.
We will ignore the possible mixing to the light generations which is strongly constrained by flavour physics.
The idea is to study tree level and radiative decays (loop induced decay into photon or gluon plus SM fermion) due to such Yukawa interactions to understand if the observation of such modes at the LHC would allow us to distinguish the different cases and/or estimate the size of the new Yukawa couplings.
In the next section, we will define the effective model, and study the possible Yukawa interactions as a function of the representation of the heavy fermion and the number of new fermions. 
In section 3 we limit ourselves to the third generation of quarks and define the allowed parameter space.
In section 4 we present the results for the tree and loop level decays and in section 5 we discuss the numerical results and briefly the LHC prospects.

\section{The effective model}

In the following we shall assume that the new fermions interact with the SM fermions via Yukawa interactions.
The quantum numbers of the new fermions with respect to the weak SU(2)$_L\times$ U(1)$_Y$ gauge group are therefore limited by the requirement of an interaction with the Higgs doublet and one of the SM fermions.
The standard model contains a doublet $q_L = \{ u_L, d_L \} = (2,Y)$ and two singlets $u_R = (1,Y+\frac{1}{2})$ and $d_R = (1,Y-\frac{1}{2})$ where $Y=\frac{1}{6}$ for quarks and $Y=\frac{1}{2}$ for leptons, and the Higgs $H = (2,\frac{1}{2})$.
The SM Yukawa couplings are:
\beq \label{eq:SMyuk}
\mathcal{L}_{\rm Yuk} = - y_u\, \bar q_L H^c u_R - y_d\, \bar q_L H d_R + h.c.\,,
\eeq
Taking into account the quantum numbers of the standard model particles one can easily check the possible quantum number assignments for the new fermions. One can add a new singlet fermion with the same hypercharge assignments 
as in the SM, namely $Y \pm \frac{1}{2}$. There are 3 possible doublets: one with the SM hypercharge $Y$, and two others with $Y \pm 1$. Finally, one can add two triplets with hypercharge $Y \pm \frac{1}{2}$.

In the following we will denote by $U$ and $D$ the heavy partners of the up and down SM particles, namely the 
states that will mix with the SM fermions. We will denote by a $X$ the eventual extra fermion that does not mix with SM ones, because of a different electric charge.

\subsection{Two fermion mixing}
\label{sec:mix2}

The Yukawa coupling $\lambda$ connecting the heavy fermions with the SM ones will generate a mixing between the two states, with the light one to be identified with the SM mass eigenstate.
In general, there are two types of mixing: the singlets and triplets couple to the left-handed doublet, while the doublets couple with the right-handed singlets.
In the following we will study these two cases in general, adding two heavy states, $U$ and $D$, and parametrising their mixing with the SM states.
This formalism can then easily adapted to the different representations of the heavy fermions.

In the case of singlets and triplets, after the Higgs doublet develops a vacuum expectation value
\beq
\langle H \rangle = \left( \begin{array}{c}
0 \\ \frac{v + h}{\sqrt{2}} \end{array} \right)\,,
\eeq
where $v \sim 246$ GeV and $h$ is the physical Higgs boson, the mass terms will look like
\beq
\mathcal{L}_{\rm mass} = - \frac{y_u v}{\sqrt{2}}\, \bar u_L u_R - x\, \bar u_L U_R - M\, \bar U_L U_R + h.c.\, ,
\label{eq:singl}
\eeq
where $x \sim \lambda v$ with the proportionality factor depending on the representation $U$ belongs to (a similar expression holds for down-type fermions). 
In the singlet case, a mass term $\bar U_L u_R$ is also allowed, however one can always find a combination of $U_R$ and $u_R$ to remove such parameter and redefine the Yukawa couplings.

\begin{figure}[tb]
\begin{center}
\epsfig{file=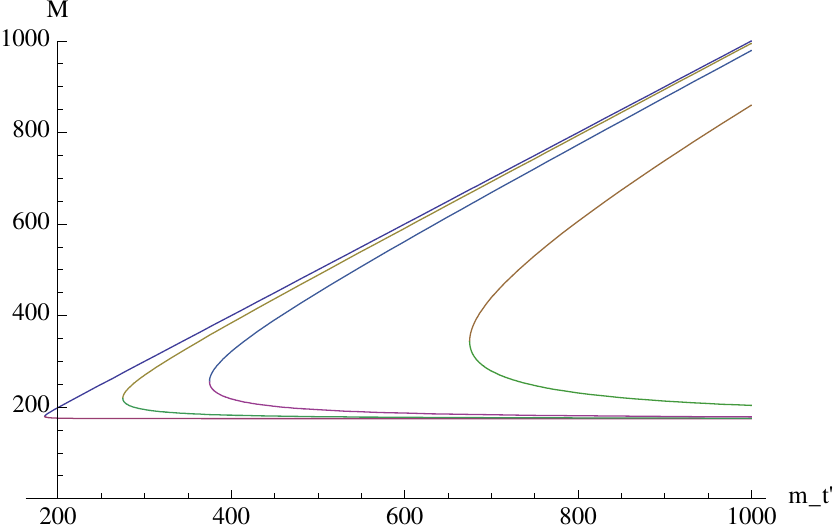,width=.7\textwidth}
\caption{\sl Plot of $M$ as a function of $m_{t^\prime}$ for different values of $x$ (from left to right: $10$, $100$, $200$ and $500$ GeV). In the limit case $x\to 0$, the two straight lines correspond to $M = m_{t^\prime}$, and $M = m_t$ (with $m_{t^\prime} = \frac{y_u v}{\sqrt{2}}$).}
\label{fig:M}
\end{center}
\end{figure}

The mass matrix can be diagonalised by two mixing matrices
\beq
V_{u}^{L,R} = \left( \begin{array}{cc}
\cos \theta_u^{L,R} & \sin \theta_u^{L,R} \\
- \sin \theta_u^{L,R} & \cos \theta_u^{L,R}
\end{array} \right)\,,
\eeq
defined as
\begin{eqnarray}
\left(
\begin{array}{cc}
\cos \theta^{L}_{u} & -\sin \theta^{L}_{u} \\
\sin \theta^{L}_{u} & \cos \theta^{L}_{u} \\
\end{array}
\right)
\left(
\begin{array}{cc}
\frac{y_u v}{\sqrt{2}} & x \\
0 & M \\
\end{array}
\right)
\left(
\begin{array}{cc}
\cos \theta^{R}_{u} & \sin \theta^{R}_{u} \\
-\sin \theta^{R}_{u} & \cos \theta^{R}_{u} \\
\end{array}
\right) =
\left(
\begin{array}{cc}
m_{t} & 0 \\
0 & m_{t^{\prime}} \\
\end{array}
\right)\, ,
\end{eqnarray}
where  $m_{t^{\prime}} \geq M \geq m_{t}$.
The relations between the three input parameters and the mixing angles and masses can be expressed as
\begin{eqnarray}
\frac{y_u^2 v^2}{2} &=& m_{t}^{2} \left(1 + \frac{x^{2}}{M^{2} - m_{t}^{2}} \right)\, , \\
m_{t^{\prime}}^{2} &=& M^{2} \left(1 + \frac{x^{2}}{M^{2} - m_{t}^{2}} \right)\, ,\\
\sin \theta^{L}_{u} &=& \frac{M x}{\sqrt{(M^{2} - m_{t}^{2})^{2} + M^{2}x^{2}}}\, , \\
\sin \theta^{R}_{u} &=& \frac{m_t}{M} \sin \theta^{L}_{u}\, .
\end{eqnarray}
In this case, for $M \gg m_t$, the right-handed mixing angle is much smaller that the left-handed one.
Note also that some values of $m_{t^\prime}$ can be obtained by two choices of $M$ (see Figure~\ref{fig:M}): for large $M$ the top mostly lives in the chiral fermion, while for the small $M$ solution it mostly consists of the new fermion (and the Yukawa $y_u$ is large).
For $M \gg v \sim m_t \sim x$:
\beq
\frac{y_u v}{\sqrt{2}} \sim m_t\, , & \quad & m_{t^\prime} \sim M\, ; \nonumber\\
\sin \theta^{L}_{u} \sim \frac{x}{M}\, , & \quad &\sin \theta^{R}_{u} \sim \frac{ m_t x}{M^2}\, .
\eeq 

In the case of doublets:
\beq
\mathcal{L}_{\rm mass} = - \frac{y_u v}{\sqrt{2}}\, \bar u_L u_R - x\, \bar U_L u_R - M\, \bar U_L U_R + h.c.\, .
\eeq
The diagonalisation can be calculated in a similar way:
\begin{eqnarray}
\left(
\begin{array}{cc}
\cos \theta^{L}_{u} & -\sin \theta^{L}_{u} \\
\sin \theta^{L}_{u} & \cos \theta^{L}_{u} \\
\end{array}
\right)
\left(
\begin{array}{cc}
\frac{y_u v}{\sqrt{2}} & 0 \\
x & M \\
\end{array}
\right)
\left(
\begin{array}{cc}
\cos \theta^{R}_{u} & \sin \theta^{R}_{u} \\
-\sin \theta^{R}_{u} & \cos \theta^{R}_{u} \\
\end{array}
\right) =
\left(
\begin{array}{cc}
m_{t} & 0 \\
0 & m_{t^{\prime}} \\
\end{array}
\right)\, ;
\end{eqnarray}
the relations between parameters are the same except that the formulas for the left- and right-handed mixing angles are exchanged:
\begin{eqnarray}
\sin \theta^{R}_{u} &=& \frac{M x}{\sqrt{(M^{2} - m_{t}^{2})^{2} + M^{2}x^{2}}}\, , \\
\sin \theta^{L}_{u} &=& \frac{m_t}{M} \sin \theta^{R}_{u}\, .
\end{eqnarray}
In this case, therefore, it is the left-handed angle to be small for large $M$.

Using the mixing matrices, we can express the couplings to $Z$, $W$ and $h$ as two by two matrices in the mass eigenstate basis (the couplings with the photon and gluon stay diagonal due to gauge invariance).
If we denote by $g_W^{sm}$ and $g_W^\psi$ the couplings of the $W$ with the SM doublet and the new fermion respectively, the left-handed couplings can be written as:
\beq
g_{WL} = \left( V_{d}^L \right)^\dagger \cdot \left( \begin{array}{cc}
g_W^{sm} & 0 \\
0 & g_{W}^\psi
\end{array} \right) \cdot V_u^L  = \left( \begin{array}{cc}
g_W^{sm} c_d^L c_u^L + g_W^\psi s_d^L s_u^L & g_W^{sm} c_d^L s_u^L - g_W^\psi s_d^L c_u^L \\
g_W^{sm} s_d^L c_u^L - g_W^\psi c_d^L s_u^L  & g_W^{sm} s_d^L s_u^L + g_W^\psi c_d^L c_u^L 
\end{array} \right)\,,
\eeq
where $s$ and $c$ stand for the $\sin$ and $\cos$ of the mixing angles.
The same formula applies for the right-handed couplings, with $g_W^{sm} = 0$:
\beq
g_{WR} = \left( V_{d}^R \right)^\dagger \cdot \left( \begin{array}{cc}
0 & 0 \\
0 & g_{W}^\psi
\end{array} \right) \cdot V_u^R = \left( \begin{array}{cc}
 g_W^\psi s_d^R s_u^R & - g_W^\psi s_d^R c_u^R \\
 - g_W^\psi c_d^R s_u^R  & g_W^\psi c_d^R c_u^R 
\end{array} \right)\,.
\eeq
Note that $g_W^{sm} = \frac{g}{\sqrt{2}}$, and $g_W^\psi$, the same for left- and right-handed components, depends on the representation: it is equal to the SM one for a doublet and equal to $\pm g$ for a triplet.
Note also that in the case where either $U$ or $D$ are absent, the same formulas can be used just setting $g_W^\psi = 0$ and setting to zero the absent mixing angle.
Similarly, a general matrix formula can be written for the $Z$ couplings of both left- and right-handed ups and downs:
\beq
g_{Zf} = \left( V_{f} \right)^\dagger \cdot \left( \begin{array}{cc}
g_Z^{sm} & 0 \\
0 & g_Z^\psi
\end{array} \right) \cdot V_f = \left( \begin{array}{cc}
g_Z^{sm} c^2 + g_Z^\psi s^2 & (g_Z^{sm}  - g_Z^\psi ) s c \\
(g_Z^{sm} - g_Z^\psi )  s c  & g_Z^{sm} s^2 + g_Z^\psi c^2 
\end{array} \right)\,.
\eeq
Note here that the $Z$ couplings can be always expressed as function of the weak isospin and charge of the fermion:
\beq
g_Z (T_3, Y) = \frac{g}{\cos \theta_W} \left( T_3 - \sin^2 \theta_W Q \right)\,.
\eeq
Finally, we can express the Higgs couplings in the two cases as:
\beq
\lambda_h = \left( V^L \right)^\dagger \cdot \left( \begin{array}{cc} \frac{y}{\sqrt{2}} & \frac{x}{v} \\ 0 & 0 \end{array} \right) \cdot V^R &=& \left( \begin{array}{cc}
c^L \left( c^R \frac{y}{\sqrt{2}} - s^R \frac{x}{v} \right) & c^L \left(s^R \frac{y}{\sqrt{2}} + c^R \frac{x}{v}\right) \\
s^L \left(c^R \frac{y}{\sqrt{2}} - s^R \frac{x}{v} \right) & s^L \left(s^R \frac{y}{\sqrt{2}} + c^R \frac{x}{v}\right)
\end{array} \right)\,, \\
\lambda_h = \left( V^L \right)^\dagger \cdot \left( \begin{array}{cc} \frac{y}{\sqrt{2}} & 0 \\ \frac{x}{v} & 0 \end{array} \right) \cdot  V^R  &=& \left( \begin{array}{cc}
c^R \left(c^L \frac{y}{\sqrt{2}} - s^L \frac{x}{v}\right) & s^R \left(c^L \frac{y}{\sqrt{2}} - s^L \frac{x}{v}\right) \\ 
c^R \left(s^L \frac{y}{\sqrt{2}} + c^L \frac{x}{v}\right) & s^R \left(s^L \frac{y}{\sqrt{2}} + c^L \frac{x}{v}\right)
\end{array} \right)\,.
\eeq

\section{Third generation quarks}
\setcounter{equation}{0}
\setcounter{footnote}{0}

Precision tests on new vector-like fermions play an important role to constrain the possible range of masses, couplings and mixings. 
For light quarks and leptons, the modification of tree level couplings to the gauge bosons poses a very strong bound on the masses or mixing angles, therefore we will ignore this case.
The only case that may be relevant for the LHC is a mixing in the third generation of quarks: in the following we will consider only mixing involving top and bottom quarks, and assume that the mixing effects are negligible for the light generations so that they do not appear in flavour measurements.

Nevertheless, tree level bounds, typically coming from $W \to tb$ or $Z \to b\bar b$ due to mixing effects with the new heavy fermions, give important indications: they are ``robust'' bounds,  in the sense that they only depend on the mixing parameter and the properties of the new particle.
Loop-level bounds, as for example oblique corrections, are also an important and often stringent test of the possible range of parameters for new heavy fermions, however they are more model dependent than the previous ones.
In fact, heavier particles that appear in a specific model also contribute leading to potential cancellations.
Moreover, the Higgs mass is still unmeasured and its variation from the reference value can compensate for the heavy fermion contributions. 
The importance of the oblique corrections is due to the fact that in many extensions of the Standard Model the vacuum-polarisation diagrams are the main corrections to the standard particle interactions. 
However the oblique corrections are not much useful if the new particles mix strongly with the standard ones as in this case direct corrections to the standard particle will be much more important \cite{quarkmixing}. We will consider both type of direct and oblique corrections in the following.

The direct bound to the coupling of the $W$ to top and bottom is coming from the observation of single top production at TeVatron: we will allow a variation of $\pm 20\%$ \cite{Wtbdirect}.
A tighter constraint originates from the unitarity of the CKM mixing matrix and flavour physics: however such bound is not applicable here because it does not take into account the effect of the heavy fermions, and it is also sensible to the contribution of other particles in the model (see \cite{AguilarSaavedra:2002kr} for a detailed study of bounds for singlet quarks and 
\cite{Alwall:2006bx} for a vector-like up-type quark or a fourth generation).
The couplings of the $Z$ to the bottom are also directly measured and very constrained\cite{Zbb}: in the left-handed coupling  a $+1\%$ and $-0.2\%$ deviations are allowed; in the right-handed one, $+20\%$ and $-5\%$.
In the cases under study, only one of the two is affected, so that those limits are sufficient even though the bounds are correlated; in the case where they are both present, a more detailed fit is required.

For the oblique corrections, we calculated the contribution to the $T$ parameter \cite{Peskin:1991sw}. 
A detailed study is given in \cite{Lavoura:1992np,Cynolter:2008ea}.
We allow for a deviation of $+ 0.4$ and $-0.2$: we consider a tighter bound on negative values because it is generically more difficult to accommodate for a negative shift in $T$.
For instance, increasing the Higgs mass with respect to the reference value will generate an effective negative contribution.
This is a very conservative bound and we use it just to underline the power of oblique constraints with respect to the tree level ones.
As mentioned above, model dependent contribution from other heavier particles may be relevant and give rise to cancellation, therefore significantly modifying the allowed parameter space.

Another important bound on the parameter space comes from direct searches at the TeVatron.
The most recent bounds are $335$ GeV for a $t'$ state~\cite{Tevatrontp}, and $385$ GeV for a $b'$\cite{Tevatronbp}: however, those bounds assume 100\% branching ratios $t' \to W q$ and $b' \to W t$.
This is true for a fourth generation, but in our case decays in neutral bosons will play an important role.
Case by case, we will set a bound roughly at the $Z t$ or $h t$ thresholds, where the branching ratio in $W$'s will become much smaller than 100\% .

\subsection{Case I: singlets}

\begin{figure}[tb]
\begin{center}
\epsfig{file=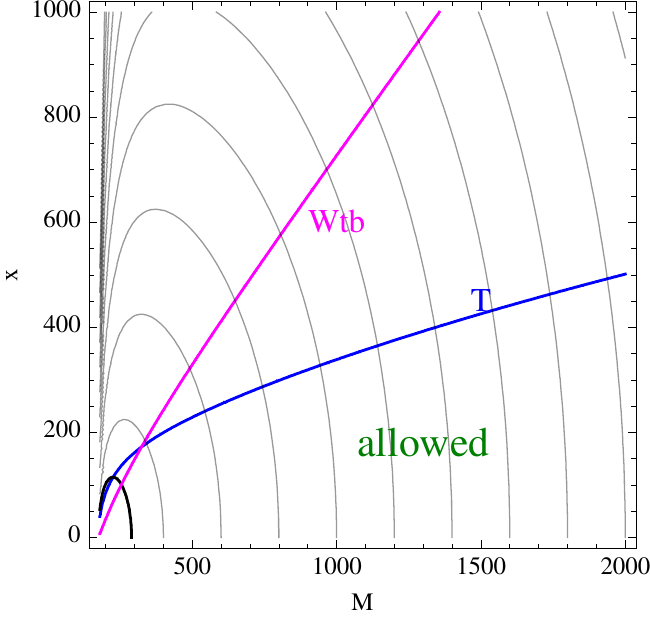,width=.46\textwidth} \epsfig{file=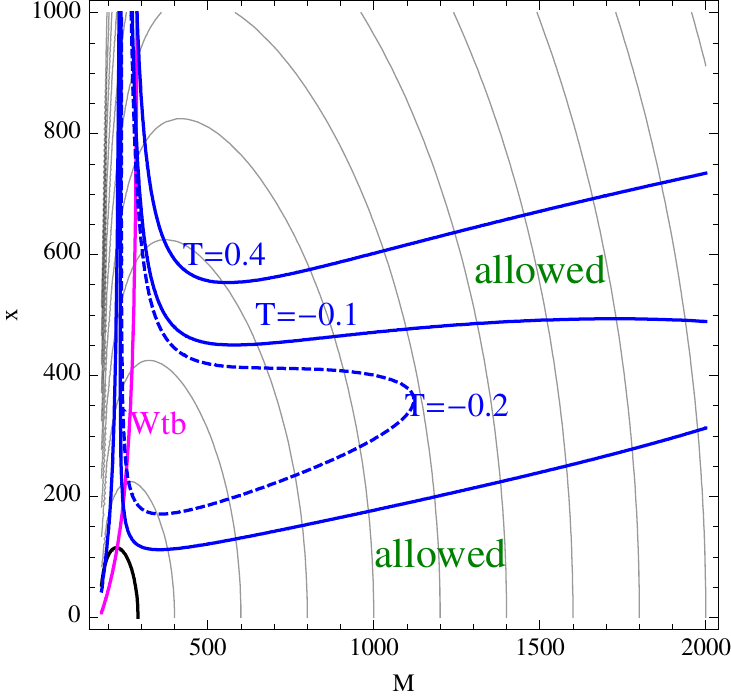,width=.46\textwidth}
\caption{\sl Singlet (left) and non-SM doublet (right) cases: in magenta the bound from $Wtb$, in blue from the $T$ parameter, in black the direct exclusion limit from TeVatron. The grey lines mark constant values of the $m_{t^\prime}$ mass (the value can be read from the intersection with the $x=0$ axis).}
\label{fig:caseI}
\end{center}
\end{figure}

Let us first consider the case $\psi = (1, \frac{2}{3} ) = U$: only a top partner is present.
The Yukawa couplings can be written as:
\beq
\mathcal{L}_{\rm Yuk} & = & - y_u\, \bar q_L H^c u_R - \lambda \, \bar q_L H^c U_R - M\, \bar U_L U_R + h.c. \nonumber \\
  & = & - \frac{y_u v}{\sqrt{2}}\, \bar u_L u_R - \frac{\lambda v}{\sqrt{2}}\, \bar u_L U_R - M\,   \bar U_L U_R + h.c. \,.
\eeq
In this case $x = \frac{\lambda v}{\sqrt{2}}$: only the up mixing is present, and the right-handed angle is smaller.
As the bottom sector is unaffected, the only tree level bound comes from the $W tb$ coupling which is reduced compared to the SM value:
\beq
\frac{\delta g_W}{g_W^{sm}} = \cos \theta_u^L - 1 = \frac{M^2-m_t^2}{\sqrt{(M^{2} - m_{t}^{2})^{2} + M^{2}x^{2}}} - 1 \sim - \frac{1}{2} \frac{x^2}{M^2}\,.
\eeq
The tree level bounds, therefore, are not very tight.
The $T$ parameter receives a positive contribution.
In Figure~\ref{fig:caseI} we show the bounds in the parameter space $x$--$M$.
For light $t^\prime$ the dominant constraint is coming from the $W$ coupling, while for heavier ones the tighter bound comes from the oblique parameters.
Changing the reference value for $T$ does not affect the blue line much, so the bound is robust.

The only heavy fermion is a top partner $t^\prime$: in this case, there are no off-diagonal right handed couplings for the $Z$ and $W$.
Therefore, it can decay to $Z t$ and $W b$ via left-handed couplings of strength respectively $\frac{1}{2} \frac{g}{\cos \theta_W} c_u^L s_u^L$ and $\frac{g}{\sqrt{2}} s_u^L$.
If kinematically allowed, the decay to $h t$ can also occur due to the new Yukawa coupling.
The branching ratios will be discussed in detail in the next sections.

In the case  $\psi = (1, -\frac{1}{3} ) = D$, only a bottom partner is present: this case is much more constrained than the previous one because the couplings of the bottom are much better measured.
The strongest tree level bound comes from corrections to $Z \bar{b}_l b_l$ coupling: 
\beq
\frac{\delta g_{ZbL}}{g_{ZbL}^{sm}} = - \frac{(s_u^L)^2}{1 - \frac{2}{3} \sin^2\theta_W}   \sim   - \frac{1}{1 - \frac{2}{3} \sin^2\theta_W} \frac{x^2}{M^2}\,.
\eeq
A negative shift in the left-handed coupling is bound to be less than about $0.2\%$:
\beq
M > 24\, x\,.
\eeq
A detailed analysis of vector-like singles effects in flavour physics is in \cite{Picek:2008dd,Arnold:2010vs}.
\subsection{Case II: SM doublet}

\begin{figure}[tb]
\begin{center}
\epsfig{file=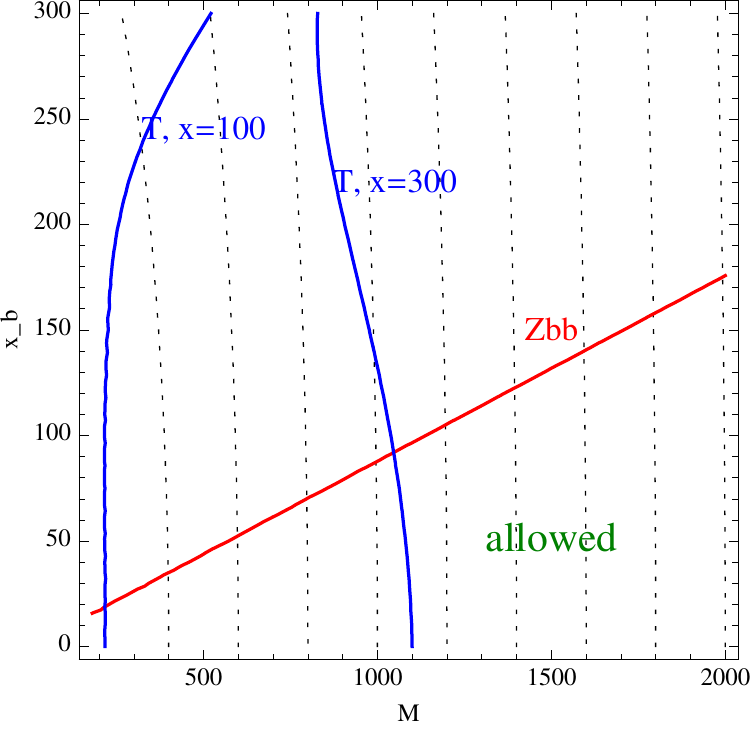,width=.46\textwidth}\,
\epsfig{file=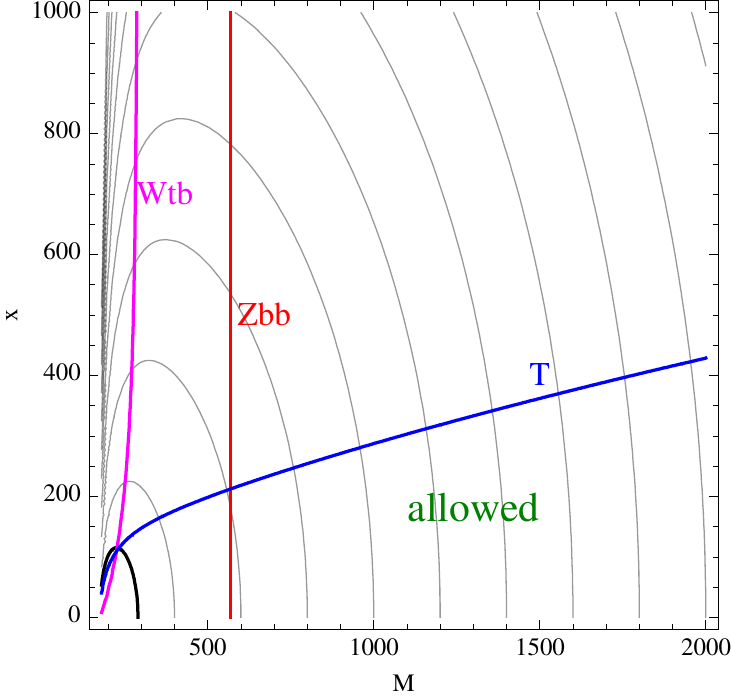,width=.46\textwidth} 
\caption{\sl SM doublet case: in magenta the bound from $Wtb$, in blue from the $T$ parameter, in red from $Zbb$, in black the direct bound from TeVatron. The grey lines mark constant values of the $m_{b^\prime}$ (left) and $m_{t^\prime}$ (right) mass. In the right panel we fixed $x_b = 50$ GeV, however the bounds from $T$ and $Wtb$ are not very sensitive to the value.}
\label{fig:caseII}
\end{center}
\end{figure}

In the SM doublet case $\psi = ( 2, \frac{1}{6} ) = \{U, D\}^T$ the vector-like fermion contains  both a top and bottom partners.
Like in the SM, it is possible to write two Yukawa couplings involving the left-handed components of the doublet:
\beq
\mathcal{L}_{\rm Yuk} & = & - y_u\, \bar q_L H^c u_R - \lambda_u \, \bar \psi_L H^c u_R  - \lambda_d \, \bar \psi_L H d_R - M\, \bar \psi_L \psi_R + h.c. \nonumber \\
  & = & - \frac{y_u v}{\sqrt{2}}\, \bar u_L u_R - \frac{\lambda_u v}{\sqrt{2}}\, \bar U_L u_R - \frac{\lambda_d v}{\sqrt{2}}\, \bar D_L d_R - M\,  ( \bar U_L U_R + \bar D_L D_R  )+ h.c. \,.
\eeq
Both up and down mixings are present and we can define $x = \frac{\lambda_u v}{\sqrt{2}}$ and  $x_b = \frac{\lambda_d v}{\sqrt{2}}$: the left-handed angle is smaller.
The most dangerous bound comes from deviations in the couplings of the $Z$ to the bottom, which are sensitive to the $\lambda_b$ Yukawa coupling:
\beq
\frac{\delta g_{ZbR}}{g_{ZbR}^{sm}} = -  \frac{3 \sin^2 \theta_d^R}{2 \sin^2 \theta_W} \sim -\frac{3}{2 \sin^2 \theta_W} \frac{x_b^2}{M^2}\,,
\eeq
while the left-handed one is unaffected because $D$ has the same coupling as the left-handed SM down quark.

The only tree level bound on $x$ comes from the left-handed $W tb$ coupling:
\beq
\frac{\delta g_{WL}}{g_W^{sm}} = \cos \theta_u^L \cos \theta_d^L +  \sin \theta_u^L \sin \theta_d^L- 1  \sim - \frac{1}{2} \frac{x^2 m_t^2}{M^4}\,.
\eeq
Due to the extra $m_t^2$ suppression, the tree level bounds are mild.
There is also a new right-handed coupling to the $W$:
\beq
\frac{g_{WR}}{g_W^{sm}} =  \sin \theta_u^R \sin \theta_d^R  \sim \frac{x x_b}{M^2}\,.
\eeq

The $T$ parameter receives a positive contribution and it depends on both $x$ and $x_b$.
From Figure~\ref{fig:caseII} we can see that the bound on $x_b$ is dominated by the $Z$ coupling: for comparison we show the bounds from $T$ for $x=100$ and $300$ GeV showing the strong dependence on $x$.
The figure also shows the bounds on $x$, as a function of $M$ for $x_b = 50$. 
The bounds from $T$ and the $W$ coupling do not depend much on the precise value of $x_b$, as long as it is small ($x_b < 150$ GeV in our range of interest).
On the other hand, the bound from the $Z$ coupling (red line) does not depend on $x$ and it shifts according to the value of $x_b$.

The physical spectrum contains a top partner $t'$ and a bottom partner $b'$:  the bottom partner is typically lighter that the top one, as long as $x_b$ is small, as it can be seen in Figure~\ref{fig:caseII}.
Therefore, the allowed decays are $b' \to (Z,h) b$ and $b' \to W t$.
On the other hand, $t' \to (Z,h) t$ and $t' \to W b$, with the additional channel $t' \to W b'$ open if $m_{t^\prime} - m_{b^\prime} > m_W$ (for large $x$).

\subsection{Case III: non-SM doublets}

In the case $\psi = (2, \frac{7}{6} ) = \{X, U\}^T$, the vector-like fermion contains  a top partner together with a new fermion $X$ with charge $\frac{5}{3}$.
The Yukawa couplings involve the left-handed component of $\psi$:
\beq
\mathcal{L}_{\rm Yuk} & = & - y_u\, \bar q_L H^c u_R - \lambda \, \bar \psi_L H u_R - M\, \bar \psi_L \psi_R + h.c. \nonumber \\
  & = & - \frac{y_u v}{\sqrt{2}}\, \bar u_L u_R - \frac{\lambda v}{\sqrt{2}}\, \bar U_L u_R - M\,  ( \bar U_L U_R + \bar X_L X_R  )+ h.c. \,.
\eeq
In this case $x = \frac{\lambda v}{\sqrt{2}}$: only the up mixing is present, and the left-handed angle is smaller.
The only tree level bound comes from the left-handed $W tb$ coupling:
\beq
\frac{\delta g_W}{g_W^{sm}} = \cos \theta_u^L - 1  \sim - \frac{1}{2} \frac{x^2 m_t^2}{M^4}\,.
\eeq
Due to the extra $m_t^2$ suppression, the tree level bounds are negligible.
The $T$ parameter can receive both a positive and a negative contribution.
For positive $T$ we fix the bound at $0.4$, and the curve does not depend much on the precise value (solid blue line in Figure~\ref{fig:caseI}).
For negative contributions, we impose a tighter bound at $-0.2$, the reason being that it is generically more difficult to accommodate for a negative shift in $T$.
In the latter case, the curve is very sensitive to the precise value (blue dashed line in Figure~\ref{fig:caseI}) and two fine tuned regions on the small $M$ branch are still allowed.

The physical spectrum contains a top partner $t'$ and a lighter new fermion $X$ with charge $\frac{5}{3}$ and mass $m_X = M$.
The only decay channel for $X$ is into $W^+ t$, where the $W$ is virtual if $M < m_t + m_W$. 
If $m_{t^\prime} - m_X > m_W$, then $t' \to W^+ X$ mostly, with a sub-leading channel in $t' \to (Z,h) t$.
The $t' \to W^+ b$ channel is suppressed by an extra power of $v/M$ in the coupling.
If $ m_{t^\prime} - m_X < m_W$, then $t' \to (Z,h) t$ is the main channel with a small contribution to $t' \to W^+ b$.

In the case  $\psi = (2, -\frac{5}{6} ) = \{D, X\}^T$, a bottom partner and a fermion $X$ with charge $- \frac{4}{3}$ are present, and a tighter constraint comes from deviation of the $Z$ couplings with the bottom.
The strongest tree level bound comes from corrections to the $Z \bar{b} b$ coupling: 
\beq
\frac{\delta g_{ZbL}}{g_{ZbL}^{sm}} &=&  -\frac{2}{1-\frac{2}{3} \sin^2 \theta_W} \sin^2 \theta_d^L \sim  -\frac{2}{1-\frac{2}{3} \sin^2 \theta_W} \frac{x^2 m_b^2}{M^4}\,, \\
\frac{\delta g_{ZbR}}{g_{ZbR}^{sm}} &=&  \frac{3 \sin^2 \theta_d^R}{2 \sin^2 \theta_W} \sim \frac{3}{2 \sin^2 \theta_W} \frac{x^2}{M^2}\,.
\eeq
The tree level decays are similar to the other doublet, replacing top with bottom.

\subsection{Case IVa: triplet $\frac{2}{3}$}

\begin{figure}[tb]
\begin{center}
\epsfig{file=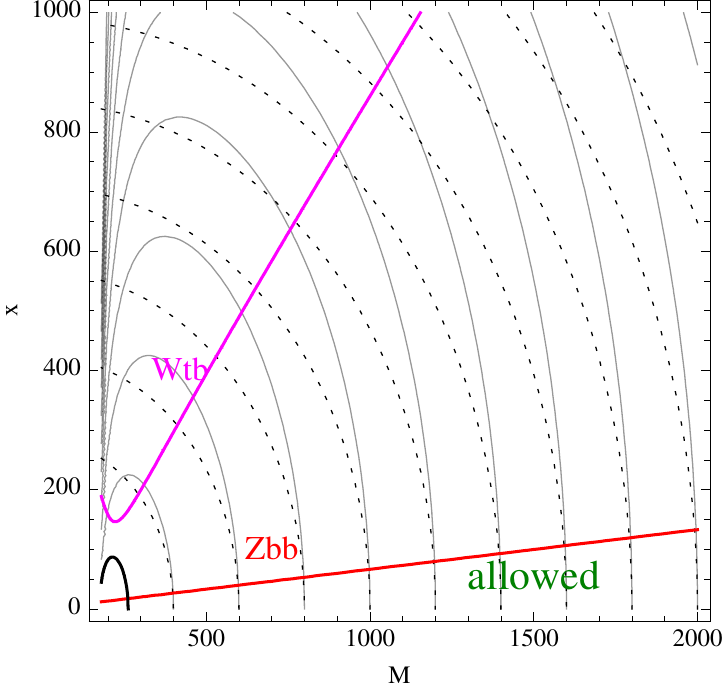,width=.46\textwidth}
\epsfig{file=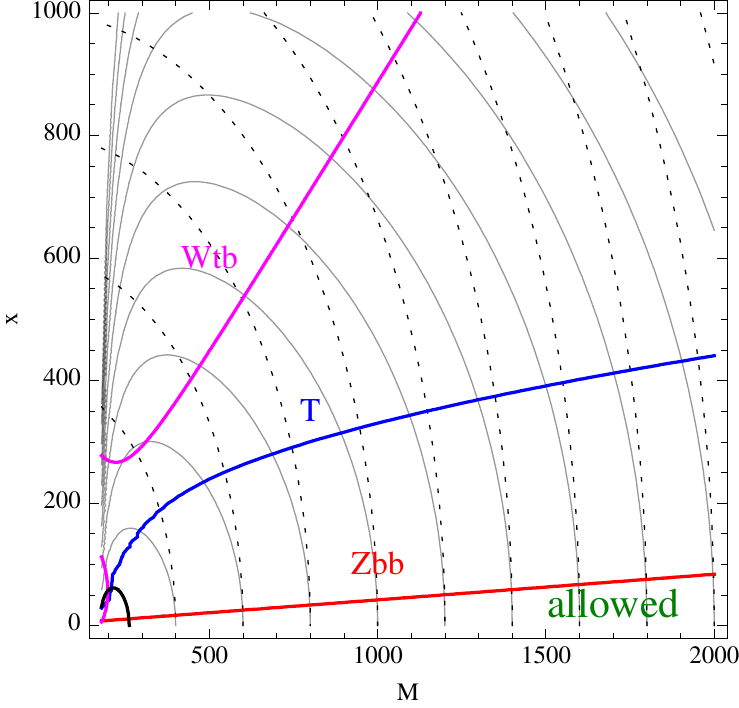,width=.46\textwidth}
\caption{\sl Triplet cases: $Y=2/3$ left, $Y=-1/3$ right. In magenta the bound from $Wtb$, in red from the $Zbb$, in blue the $T$ parameter and in black the direct bound from TeVatron. The grey (dashed) lines mark constant values of the $m_{t^\prime}$ ($m_{b^\prime}$) mass (the value can be read from the intersection with the $x=0$ axis).}
\label{fig:caseIV}
\end{center}
\end{figure}

In the case $\psi = (3, \frac{2}{3}) = \{X,U,D\}^T$, the new fermion contains a partner for both top and bottom, plus $X$ with charge $5/3$.
The Yukawa couplings involve the right-handed component of $\psi$:
\beq
\mathcal{L}_{\rm Yuk} & = & - y_u\, \bar q_L H^c u_R - \lambda \, \bar q_L \tau^a H^c \psi^a_R - M\, \bar \psi_L \psi_R + h.c. \\
  & = & - \frac{y_u v}{\sqrt{2}}\, \bar u_L u_R - \frac{\lambda v}{\sqrt{2}}\, \bar u_L U_R - \lambda v\, \bar d_L D_R - M\,  ( \bar U_L U_R + \bar D_L D_R + \bar X_L X_R  )+ h.c.\,.  \nonumber
\eeq
The new Yukawa, therefore, induces a mixing both in the top and bottom sector: defining as usual $x=\frac{\lambda v}{\sqrt{2}}$, for the top the same formulas as in case I apply, while for the bottom it suffices to replace $x \to \sqrt{2} x$.
From Figure~\ref{fig:caseIV} we can see that in most of the parameter space, especially on the large $M$ branch, $m_{b^\prime} > m_{t^\prime} > m_X$.

The presence of a bottom partner induces deviations in the couplings to the $Z$:
\beq
\frac{\delta g_{ZbL}}{g_{ZbL}^{sm}} &=&  \frac{1}{1-\frac{2}{3} \sin^2 \theta_W} \sin^2 \theta_d^L \sim  \frac{2}{1-\frac{2}{3} \sin^2 \theta_W} \frac{x^2}{M^2}\,, \\
\frac{\delta g_{ZbR}}{g_{ZbR}^{sm}} &=& - \frac{3 \sin^2 \theta_d^R}{\sin^2 \theta_W} \sim -\frac{6}{\sin^2 \theta_W} \frac{x^2 m_b^2}{M^4}\,.
\eeq
The strongest bound is therefore coming from the positive deviation to the left-handed coupling.
We also computed the $T$ parameter, however it is small and does not pose any further bound.
From Figure~\ref{fig:caseIV} we can see that the $Z$ coupling bound is very tight indeed and the new Yukawa $x$ has to be smaller than about $100$ GeV in the range of masses we considered.

The spectrum contains tree new particles, $X$, $t^\prime$ and $b^\prime$: the lightest one is always the $X$ ($m_X = M$) which decays in $W^+ t$ (with a virtual $W$ for very light $M$).
For the heavy top, $t^\prime \to (Z,h) t$ and $t^\prime \to W b$ are the main modes together with $t^\prime \to W X$.
For the heavy bottom,  $b^\prime \to (Z,h) b$ and $b^\prime \to W t$ are the main modes together with $b^\prime \to W t^\prime$.

\subsection{Case IVb: triplet $-\frac{1}{3}$}

In the case $\psi = (3, -\frac{1}{3}) = \{U,D,X\}^T$, the new fermion contains a partner for both top and bottom, plus $X$ with charge $-4/3$.
The Yukawa couplings involve the right-handed component of $\psi$:
\beq
\mathcal{L}_{\rm Yuk} & = & - y_u\, \bar q_L H^c u_R - \lambda \, \bar q_L \tau^a H \psi^a_R - M\, \bar \psi_L \psi_R + h.c. \\
  & = & - \frac{y_u v}{\sqrt{2}}\, \bar u_L u_R - \lambda v\, \bar u_L U_R + \frac{\lambda v}{\sqrt{2}}\, \bar d_L D_R - M\,  ( \bar U_L U_R + \bar D_L D_R + \bar X_L X_R  )+ h.c.\,.  \nonumber
\eeq
As before, the new Yukawa induces a mixing both in the top and bottom sector: defining as usual $x=\frac{\lambda v}{\sqrt{2}}$, for the top the same formulas as in case I apply with $x \to \sqrt{2} x$, while for the bottom it suffices to replace $x \to - x$.
From Figure~\ref{fig:caseIV} we can see that in the whole parameter space $m_{t^\prime} > m_{b^\prime} > m_X$.

The presence of a bottom partner induces deviations in the left-handed coupling to the $Z$:
\beq
\frac{\delta g_{ZbL}}{g_{ZbL}^{sm}} = -\frac{1}{1-\frac{2}{3} \sin^2 \theta_W} \sin^2 \theta_d^L \sim  -\frac{1}{1-\frac{2}{3} \sin^2 \theta_W} \frac{x^2}{M^2}\,.
\eeq
The strongest bound is therefore coming from the negative deviation to the left-handed coupling.
The $T$ parameter receives a positive contribution, however as it can be seen from Figure~\ref{fig:caseIV} the bound is always weaker than the tree level one. 
Like in the previous case, the new Yukawa $x$ has to be smaller than about $100$ GeV in the range of masses we considered.

The spectrum contains tree new particles, $X$, $t^\prime$ and $b^\prime$: the lightest one is always the $X$ which decays in $W^- b$.
For the heavy bottom, $b^\prime \to (Z,h) b$ and $b^\prime \to W t$ are the main modes together with $b^\prime \to W X$.
For the heavy bottom,  $t^\prime \to (Z,h) t$ and $t^\prime \to W b$ are the main modes together with $t^\prime \to W b^\prime$.

\section{Tree level and radiative decays}
\label{sec:loop}
\setcounter{equation}{0}
\setcounter{footnote}{0}

The main decay modes of the heavy fermions is via heavy gauge bosons, the $W$ and $Z$, and the Higgs boson $h$.
The width is effectively controlled by the new Yukawa coupling $\lambda$, which generates the mixing with the light fermions.
Generically, all those modes are present but the relative ratios depend crucially on the quantum numbers of the new fermion. These modes are also crucial for the observation of the new states at the LHC \cite{lhcpheno}.
At loop level, new channels are added, namely decays via emission  of a photon or gluon. Even though the branching ratio is suppressed by a loop, the cleanness of the photon mode may be important for LHC strategies.
Moreover, those loops may be enhanced in the case of large Yukawa $\lambda$ and provide a tool to measure it (note that tree level branchings are roughly independent on the size of the Yukawa).
Finally, in the case of strong coupling, if the heavy states originate from a strongly interacting sector, those modes may become comparable to the tree level decays, and a measurement the modes may give a hint of the composite nature of the heavy states.

Pushed by those motivations, we explore here the loop induced decays and calculate the branching ratios in the perturbative regime, i.e. for small Yukawa $\lambda$.
Numerically, we consider the one-loop result reliable for $x \leq 500$ GeV.
A generic matrix element that describes the decay of a heavy fermion $F$ to a light one $f$ plus a vector can be parametrised as\cite{Deshpande:1981zq}:
\beq \label{eq:matel}
\mathcal{M} (F \to f V) = \bar f (q_1) \left[ (a_1 p^\mu + a_2 q_2^\mu + a_3 \gamma^\mu ) L +  (b_1 p^\mu + b_2 q_2^\mu + b_3 \gamma^\mu ) R \right] F (q_2) \epsilon_\mu (p)\,,
\eeq
where $\epsilon_\mu$ is the polarisation of the vector and $p^\mu = q_2^\mu - q_1^\mu$ is its momentum, and $L$ and $R$ are the chirality projectors.
This result is valid for general decays into a gauge boson (massive or not). 
In the case of a massless one, therefore unbroken gauge symmetry, the conservation of the fermionic current ($\mathcal{M} (\epsilon_\mu = p_\mu)=0$) implies that:
\beq
a_3 &=& - \frac{1}{2} \left( m_f a_2 +m_F b_2 \right)\,, \\
b_3 &=& - \frac{1}{2} \left( m_F a_2 + m_f b_2 \right)\,,
\eeq
when the external boson is on-shell, $p^2 = 0$.
Here, $m_F$ and $m_f$ are the masses of the heavy and light fermions.
The partial width can be expressed in terms of those parameters:
\beq \label{eq:loopGamma}
\Gamma (F \to f \gamma) = \frac{\lambda^{1/2} (1,0,m_f^2/m_F^2)\; m_F}{64 \pi} \left(1- \frac{m_f^2}{m_F^2} \right)^2 \left( |m_F a_2|^2 + |m_F b_2|^2 \right)\,;
\eeq
where the phase space function $\lambda^{1/2} (x_1, x_2, x_3) = \sqrt{x_1^2 + x_2^2 + x_3^2 - 2 x_1 x_2 - 2 x_1 x_3 - 2 x_2 x_3}$.
 For the gluon, an extra colour factor of $4/3$ must be added; for a massive boson, a more general formula applies that can be found in ~\cite{Deshpande:1981zq}.
The general one loop contributions to the matrix element in eq.(\ref{eq:matel}) are listed in Appendix~\ref{sec:loops}.

The general formula can be also used to calculate the tree level decay widths into heavy gauge bosons: in this case, the left- and right-handed couplings are given by $a_3$ and $b_3$ while all the other parameters vanish.
The width is given by:
 \begin{eqnarray} \label{eq:treeGamma}
 \Gamma &&(F \to f' V) = \frac{\lambda^{1/2} (1,m_V^2/m_F^2,m_f^2/m_F^2) m_F^3}{32 \pi m_V^2}  \left( |a_3|^2+|b_3|^2 \right) \cdot \nonumber \\
&& \left[ \left( 1 - \frac{m_f^2}{m_F^2} \right)^2 -2\frac{m_V^4}{m_F^4} +  \frac{m_V^2}{m_F^2} + \frac{m_f^2 m_V^2}{m_F^4} - 12 \frac{m_f m_V^2}{m_F^3} \frac{Re\, a_3\, Re\, b_3 + Im\, a_3\, Im\, b_3}{|a_3|^2 + |b_3|^2} \right]
 \end{eqnarray}
where $Re$ and $Im$ indicate respectively the real and imaginary part.
The tree level widths for all the cases are explicitly given in Appendix~\ref{sec:widths}.

\section{Numerical results}
\label{sec:num}
\setcounter{equation}{0}
\setcounter{footnote}{0}

In this section we report numerical results for the tree and one-loop level branching ratios in the various cases.
As a reference, we fix $m_h = 120$ GeV, and we limit the plots to $m_{t'} < 1$ TeV.
For larger masses, the production cross section at the LHC is too small to lead to a clear signature (see Figure~\ref{fig:xsec}).
In all cases we will plot curves corresponding to various values of $x$ as a function of the $t'$ mass.
The point is to show the possible range of values for the branching ratios (BR) as a function of the mass.
The relative importance of the tree level decays is typically controlled by the representation of the new fermion, therefore observing those modes can help to discriminate between the different cases.
In all the plots, the continuous line corresponds to parameter space that is allowed by all constraints, while the dotted part of the lines are excluded by the $T$ parameter only and therefore may be allowed if new contributions are present.
All the numerical results are based on the formulas provided in Appendix~\ref{sec:loops} and~\ref{sec:widths}.

\begin{figure}[tb]
\begin{center}
\epsfig{file=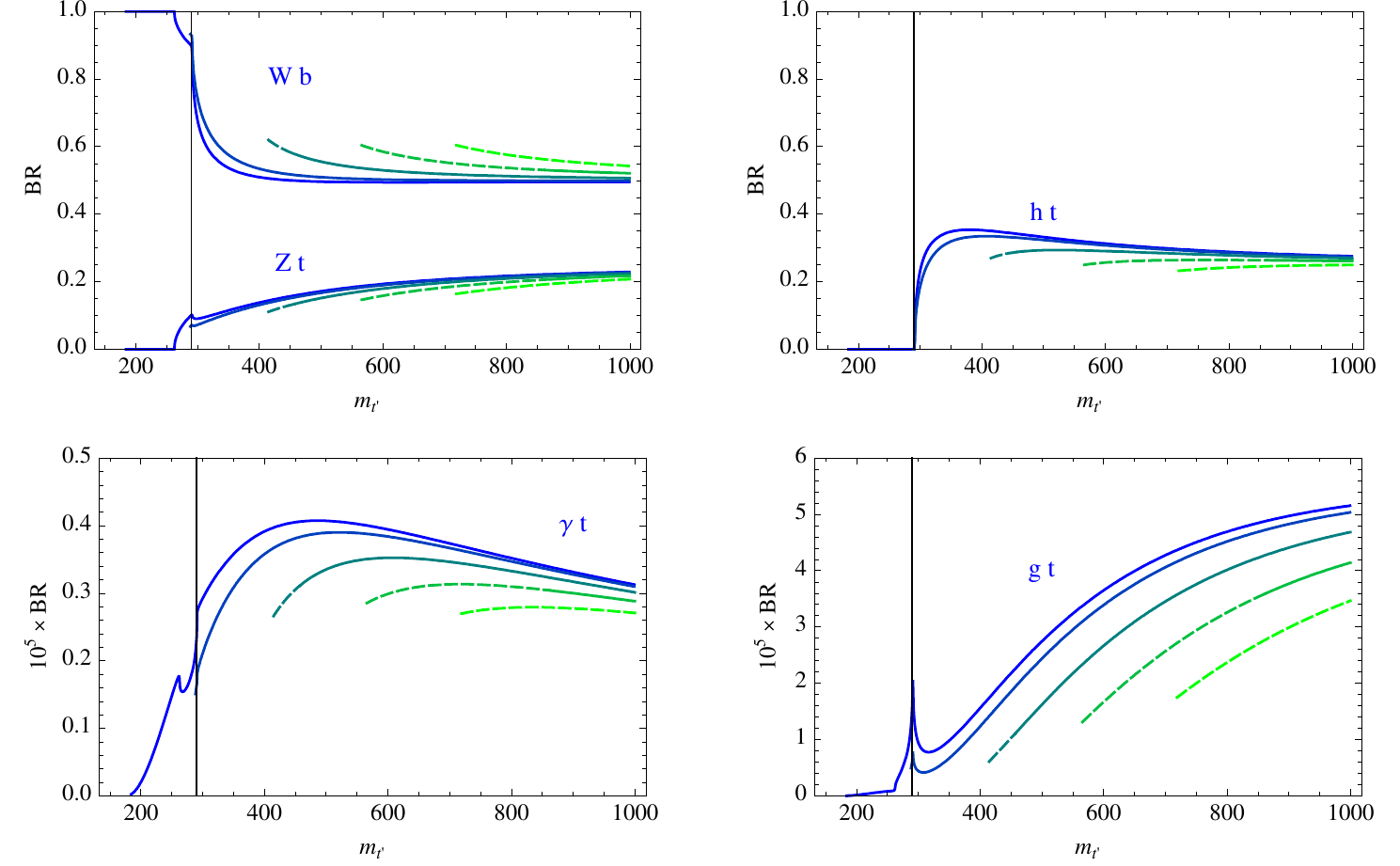,width=.97\textwidth}
\caption{\sl Singlet case: the lines correspond to $x=10, 100, 200, 300, 400$ GeV from darker (blue) to lighter gray (green); the dotted portions are excluded by the $T$ parameter. The vertical line marks the direct exclusion by TeVatron, which roughly corresponds to $290$ GeV.}
\label{fig:BR1}
\end{center}
\end{figure}

\subsubsection*{Case I: singlet}

In Figure~\ref{fig:BR1} we show the results in the singlet case for $x = 10, 100, 200, 300, 400$ GeV.
The smaller $x$ corresponds to the mostly continuous lines, while large $x$ gives rise to shorter lines due to the tighter bounds. 
Below the $h t$ threshold, the main decay is in $W b$, therefore such masses are excluded by TeVatron: the lower bound on the mass is roughly $m_h + m_t \sim 290$ GeV.
After the Higgs channel opens up, the branching ratios saturates to the values $BR (W b) \sim 50\%$ and $BR (Z t) \sim BR(h t) \sim 25\%$, in agreement with the formulae in Appendix~\ref{sec:widths}.

The loop induced channels decrease with increasing $x$ and they stay at the $10^{-6}$ level for the photonic channel and $10^{-5}$ level for the gluonic one.
We can see that those numbers agree with the rough estimates presented in Appendix~\ref{sec:estim} based on the contribution of the dominant loop.
We also observe a ``peak'' at the $h t$ threshold: this is due to the fact that below such threshold the total width decreases sharply, while the loop one increases. After the threshold, the contribution of the Higgs channel increases the total width, thus pushing down the BR.
Above threshold, we also observe a point where the BR starts decreasing due to a decrease in the partial width.

\subsubsection*{Case II: SM doublet}

\begin{figure}[tb]
\begin{center}
\epsfig{file=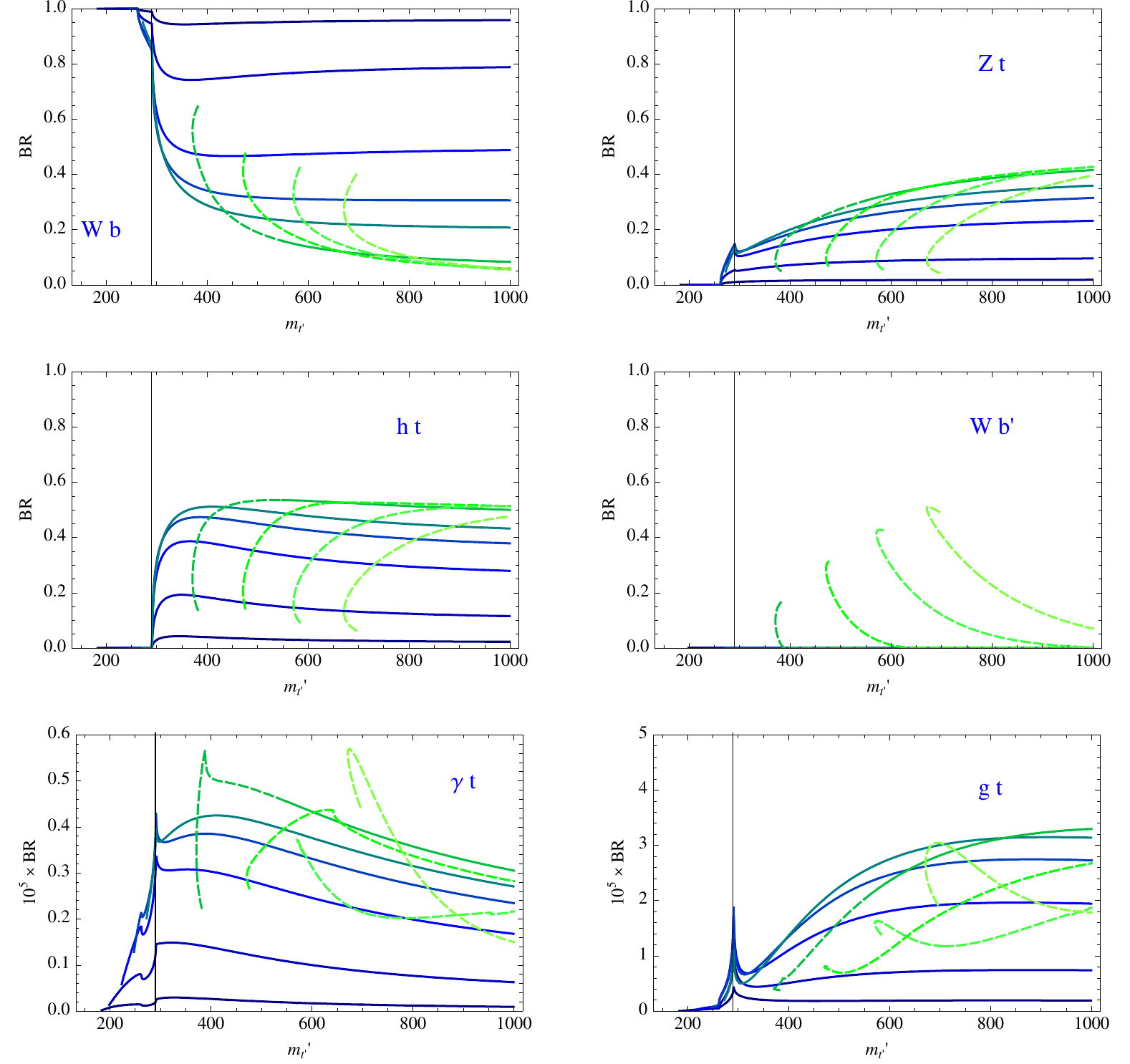,width=.97\textwidth}
\caption{\sl SM doublet: the lines correspond to $x_b = 50$ GeV and $x=10, 25, 50, 75, 100, 200, 300, 400, 500$ GeV from darker (blue) to lighter grey (green). The dotted lines are excluded by the $T$ parameter. The vertical line marks the direct exclusion by TeVatron.}
\label{fig:BR2}
\end{center}
\end{figure}

The case of the SM doublet is more complex due to the presence of two Yukawa couplings.
However, the bottom-like one is very constrained by tree level corrections, therefore here we fixed it to a small value ($x_b = 50$ GeV) and vary the top-like Yukawa $x$.
The numerical results are showed in Figure~\ref{fig:BR2}: for small $x$ the main tree level decay is $W b$, while for large values it is $(Z,h) t$.
The decay into $W b'$ is only relevant for very large $x$, values excluded by the $T$ parameter.
Also, the loop-decays tend to increase for larger values of $x$.

While the values of the BRs do not depend significantly on $x_b$, there is a lower limit on the $t'$ mass from the coupling of the $Z$ that depends crucially on the two Yukawa couplings and that can be deduced from Figure~\ref{fig:caseII}: for example, for $x_b=50$ GeV and $x=100$ GeV, $m_{t'} \geq 600$ GeV.
As for the singlet, the direct exclusion by TeVatron applies below the $h t$ threshold, where the $W b$ channel is reduced by more than 20\%.
For small $x$, the $Wb$ channel always dominates and the direct bound goes to the nominal value $335$ GeV.

\subsubsection*{Case III: non-SM doublet}

\begin{figure}[tb]
\begin{center}
\epsfig{file=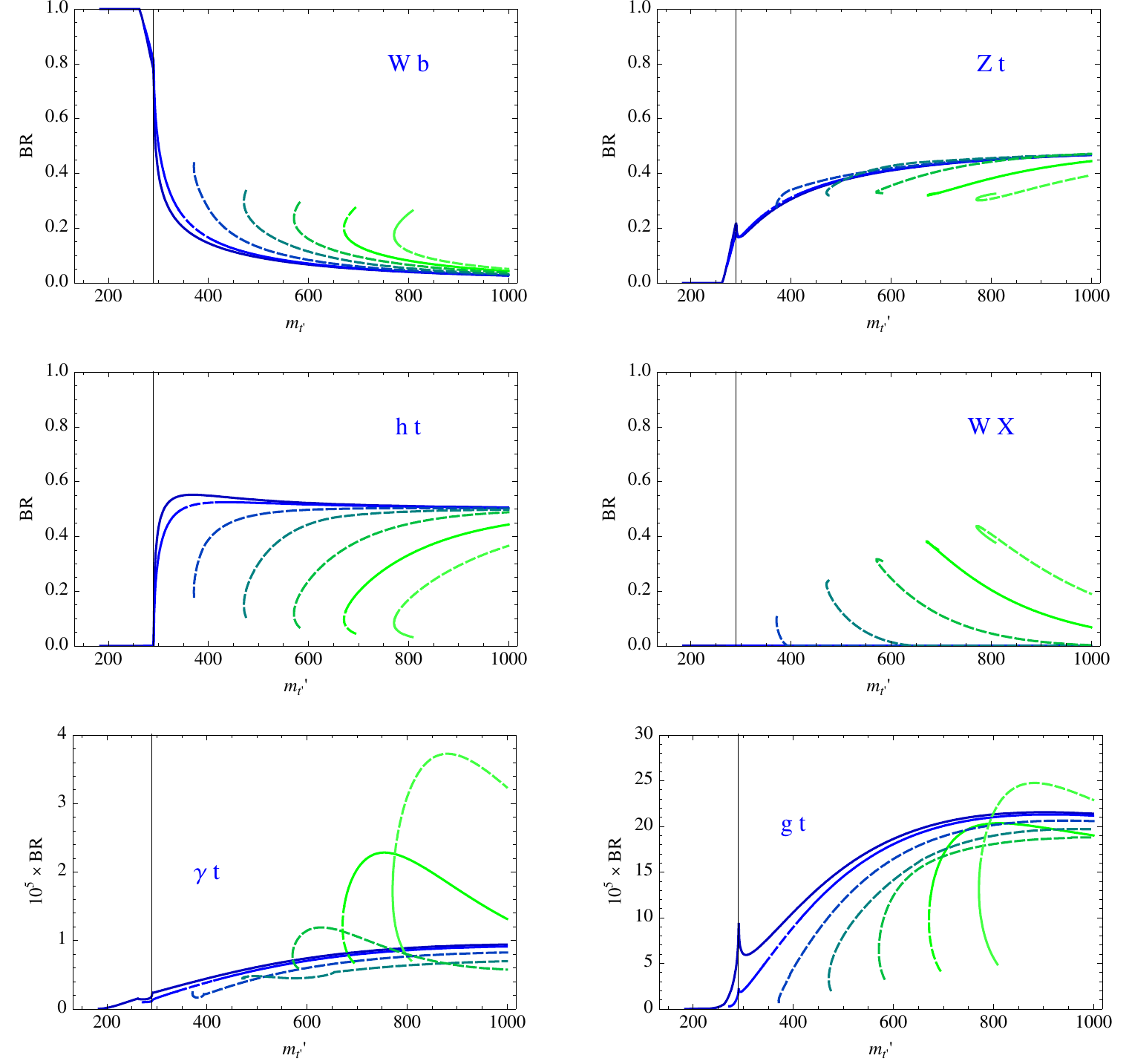,width=.97\textwidth}
\caption{\sl non-SM doublet: the lines correspond to $x=10, 100, 200, 300, 400, 500, 600$ GeV from darker (blue) to lighter grey (green). The dotted portions are excluded by the $T$ parameter. The vertical line marks the direct exclusion by TeVatron.}
\label{fig:BR3}
\end{center}
\end{figure}

In Figure~\ref{fig:BR3} we show the numerical results for the non-SM doublet case.
For small values of $x$, the tree level BR saturates rapidly to $BR(h t) \sim BR(Z t) \sim 50\%$ while the decay in $W b$ is very suppressed.
Intermediate values of $x$ are excluded by a negative contribution to $T$, while large values of $x$ are again allowed: however, $x>500$ GeV is at the edge of the non perturbative regime and the tree level and one-loop results cannot be trusted.
In this case, we also observe larger values of the loop BR compared to other cases.

\subsubsection*{Case IV: triplets}

\begin{figure}[tb]
\begin{center}
\epsfig{file=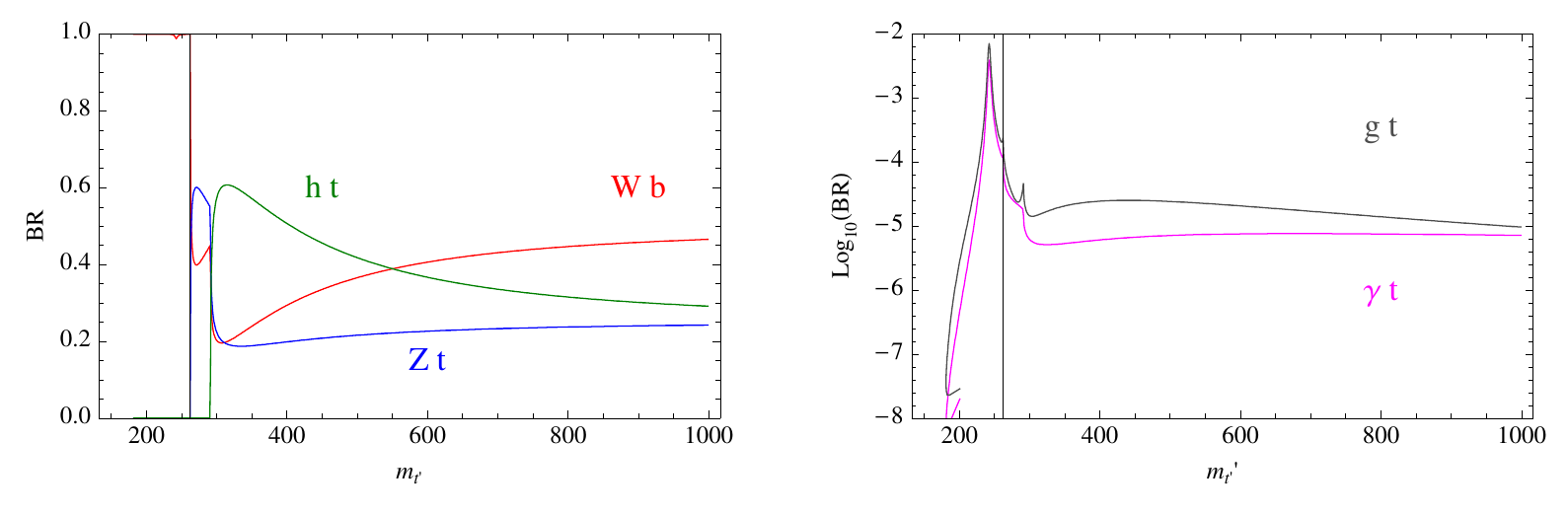,width=.97\textwidth}
\epsfig{file=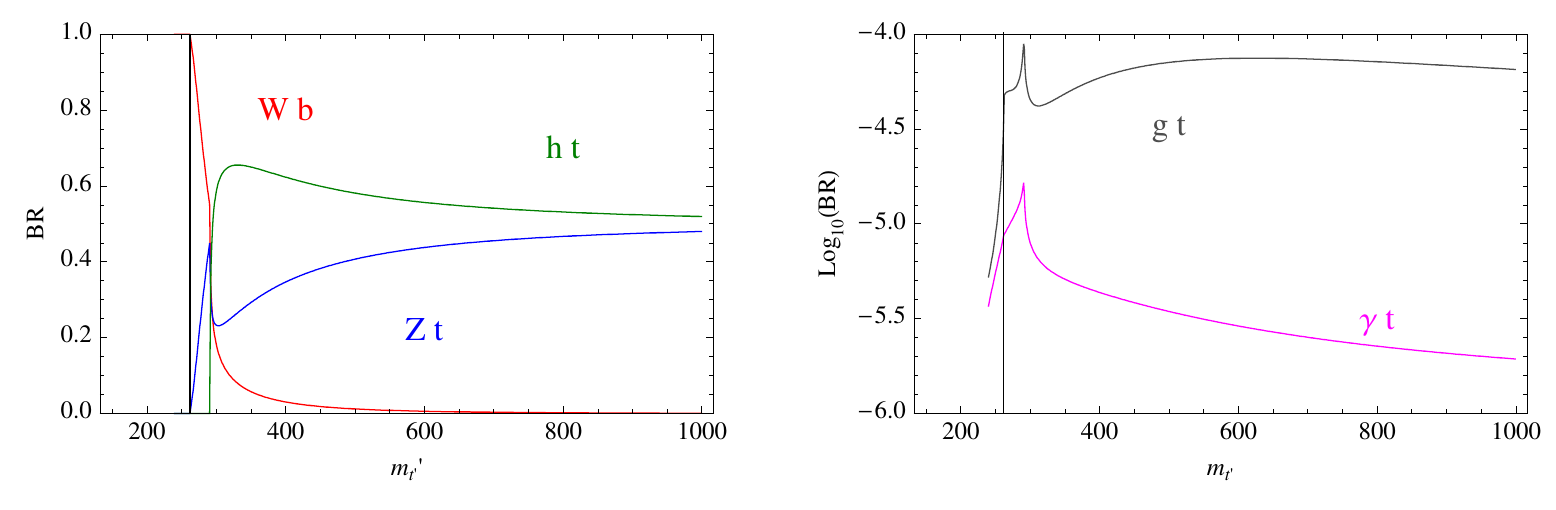,width=.97\textwidth}
\caption{\sl triplets: $Y=2/3$ top, $Y=-1/3$ bottom. The line correspond to $x=10$ GeV.  The vertical line marks the direct exclusion by TeVatron, roughly $260$ GeV.}
\label{fig:BR4}
\end{center}
\end{figure}

In Figure~\ref{fig:BR4} we show the numerical results for the two triplet cases.
As before, above the $h t$ threshold, the tree level BR saturates: while in the $Y=2/3$ case the 3 channels are comparable, in the $Y=-1/3$ case we see that the $W b$ channel is suppressed while $BR (h t) \sim BR (Z t) \sim 50\%$.
Also, the $W b$ channel starts going down after the $Z t$ threshold, so that the rough TeVatron lower bound on the mass is $m_{t'} \geq m_t + m_Z \sim 260$ GeV.
As before we observe a``peak'' at the $h t$ threshold.
In the $Y=2/3$ cases, there is also a very high peak right below the $Z t$ threshold: it results from a combination of the increase in the partial width for the radiative decays and a decrease of the total width followed by a sharp increase above the threshold.
The radiative decays can therefore reach the percent level: however this region is excluded by TeVatron because the dominant decay mode is $W b$.

\section{Conclusions}
\label{sec:concl}
\setcounter{equation}{0}
\setcounter{footnote}{0}

\begin{figure}[tb]
\begin{center}
\epsfig{file=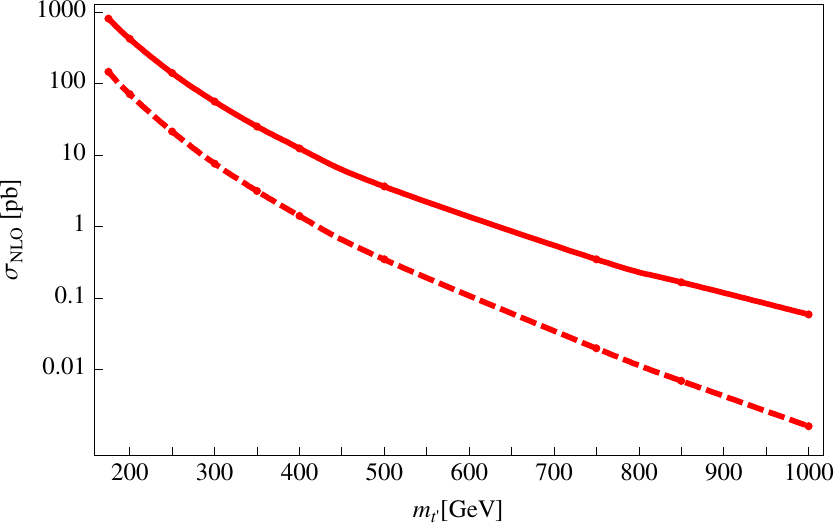,width=.6\textwidth}
\caption{\sl Heavy quarks pair production at the LHC in picobarns as a function of the heavy quark mass. The dashed (lower) line corresponds to the 7 TeV centre of mass energy while the solid (upper) line to the 14 TeV centre of mass energy case. The cross sections are normalised with a k-factor to the improved NLO values of the $\bar t t$ cross section in table 2 of \protect{\cite{Ahrens:2010mj}}.}
\label{fig:xsec}
\end{center}
\end{figure}

In this paper we have discussed the effects of new vector-like fermions, assuming that the new fermions mix with the SM ones via Yukawa interactions. Assuming a SM Higgs sector (or more generally Higgs doublets), the new heavy fermions can be singlets, doublets or triplets of isospin, with in some cases the possibility of exotic hypercharges. 
In this work, we limited ourselves to the case of only one new fermion (lighter than eventual other ones) that mixes to top and bottom only.
We have analysed present bounds at tree level and loop level, and computed the widths and branching ratios of the new particles at tree level to $W b$, $Z t$ and $h t$, and at loop level to a photon or a gluon (radiative decays). 
Our result shows that, while the tree level branching ratios depend on the mixing parameters, the radiative ones, once we limit ourselves to the allowed parameter space, are independent on the precise value of the new Yukawa coupling and close to what one would obtain with a simple estimate.
These results are very useful for model independent searches and exclusions at the LHC.
In fact, the new states will be abundantly produced both in the 7 TeV and in the 14 TeV phases of the LHC for fairly low masses.
At 7 TeV, the strong pair production cross section is above 1 pb for $m_{t'} < 600$ GeV, while at 14 TeV we have cross sections larger than 100 fb for masses below a TeV.
In Figure~\ref{fig:xsec} we plotted the estimated cross section calculated at LO for different masses of the heavy fermion and multiplied by the k-factor to the improved NLO cross section for $\bar t t$ production.
Because of the fairly large cross sections, the various tree level decay modes should be observed at the LHC.
An interesting feature is that the pattern of tree level decay modes depends on the representation of the new fermion and measuring the branching ratios in $Z t$ and $h t$ can allow to limit the number of cases and rule out the possibility of a chiral fourth generation.

Another interesting channel at the LHC is the single production of the heavy fermion, see for example 
\cite{Azuelos:2004dm,Han:2005ru}.
Even though the process is mediated by electroweak bosons and Higgs, the cross sections can be large and depend crucially on the value of the new Yukawa coupling.
However, contrary to the pair production, this channel is very model dependent and requires a more detailed study that we will present in a forthcoming publication.
The importance of this channel is that it can give direct information on the size of the new Yukawa coupling.

%

\begin{table}[tb]
\begin{center}
\begin{tabular}{|l||ccc|cc|}
\hline
case & $W b$ & $Z t$ & $h t$ & $ \gamma t$ & $g t$ \\
\hline
Singlet, $x=10$ & 0.50 & 0.17 & 0.33 & 4 $\times 10^{-6}$ & 2.7 $\times 10^{-5}$ \\
Singlet, $x=200$ & 0.50 & 0.15 & 0.29 & 3 $\times 10^{-6}$ & 1.6 $\times 10^{-5}$ \\
\hline
SM doublet, $x=10$ & 0.95 & 0.017 & 0.03 & 0.21 $\times 10^{-6}$ & 0.2 $\times 10^{-5}$ \\
SM doublet, $x=100$ & 0.24 & 0.26 & 0.50 & 4 $\times 10^{-6}$ & 2.2 $\times 10^{-5}$ \\
\hline
Non-SM doublet, $x=10$ & 0.09 & 0.37 & 0.61 & 6 $\times 10^{-6}$ & 15 $\times 10^{-5}$ \\
\hline
Triplet A (Y=2/3), $x=10$ & 0.36 & 0.22 & 0.42 & 7.2 $\times 10^{-6}$ & 2.4 $\times 10^{-5}$ \\
\hline
Triplet B (Y=-1/3), $x=10$ & 0.01 & 0.41 & 0.58 & 3.5 $\times 10^{-6}$ & 7.1 $\times 10^{-5}$ \\
\hline
\end{tabular}
\end{center}
\caption{Typical values for the branching ratios in the various cases, for $m_{t'} = 500$ GeV.} \label{tab:brs}
\end{table}

Finally we calculated the branching ratios to loop induced decays into a photon or gluon.
The branching ratios are typically of the order $10^{-6}$ for the photonic channel and $10^{-5}$ for the gluonic one in the perturbative regime, therefore they will be virtually impossible to see at the LHC. 
For illustration, in Table~\ref{tab:brs} we listed the branching ratios in the various cases for $m_{t'} = 500$ GeV.
However, those values are quite independent on the value of the Yukawa coupling, therefore observing values much larger than the ones calculated here may be a hint of strong coupling regime.
In fact, higher order operators, generated by the strong coupling, do contribute equally to the $W$, $Z$ and $\gamma$ channels: if those operators are large, we would expect order one branching ratios in photon, as typical of an excited fermion.
Larger values of the loop induced channels may also be generated by heavier states running in the loop and not directly observed at the LHC.
In this case, the loop may be enhanced by large couplings in the heavy sector.
Nevertheless, in the perturbative regime, such new contribution would be loop suppressed and therefore can not enhance the channel by orders of magnitude.
In this sense, we conclude that under our assumptions that only one ``light'' vector-like fermion is present, the observation of the $\gamma t$ or $g t$ channels at the LHC hints to a strong coupling dynamic as the origin of the new state.

\section*{Acknowledgements}
The research of Y.O. is supported in part by the
Grant-in-Aid for Science Research, Ministry of Education, Culture,
Sports, Science and Technology (MEXT), Japan, No. 16081211 and by the
Grant-in-Aid for Science Research, Japan Society for the Promotion of
Science (JSPS), No. 20244037 and No. 22244031.  This collaboration was made possible by
funding by the French ``Minist\`ere des Affaires Etrang\`eres'' and
the Japanese JSPS under a PHC-SAKURA project No. 18763QK.  

\appendix
\section*{Appendix}

\section{General loop calculation formulas}
\label{sec:loops}

\begin{figure}[tb]
\begin{center}
\epsfig{file=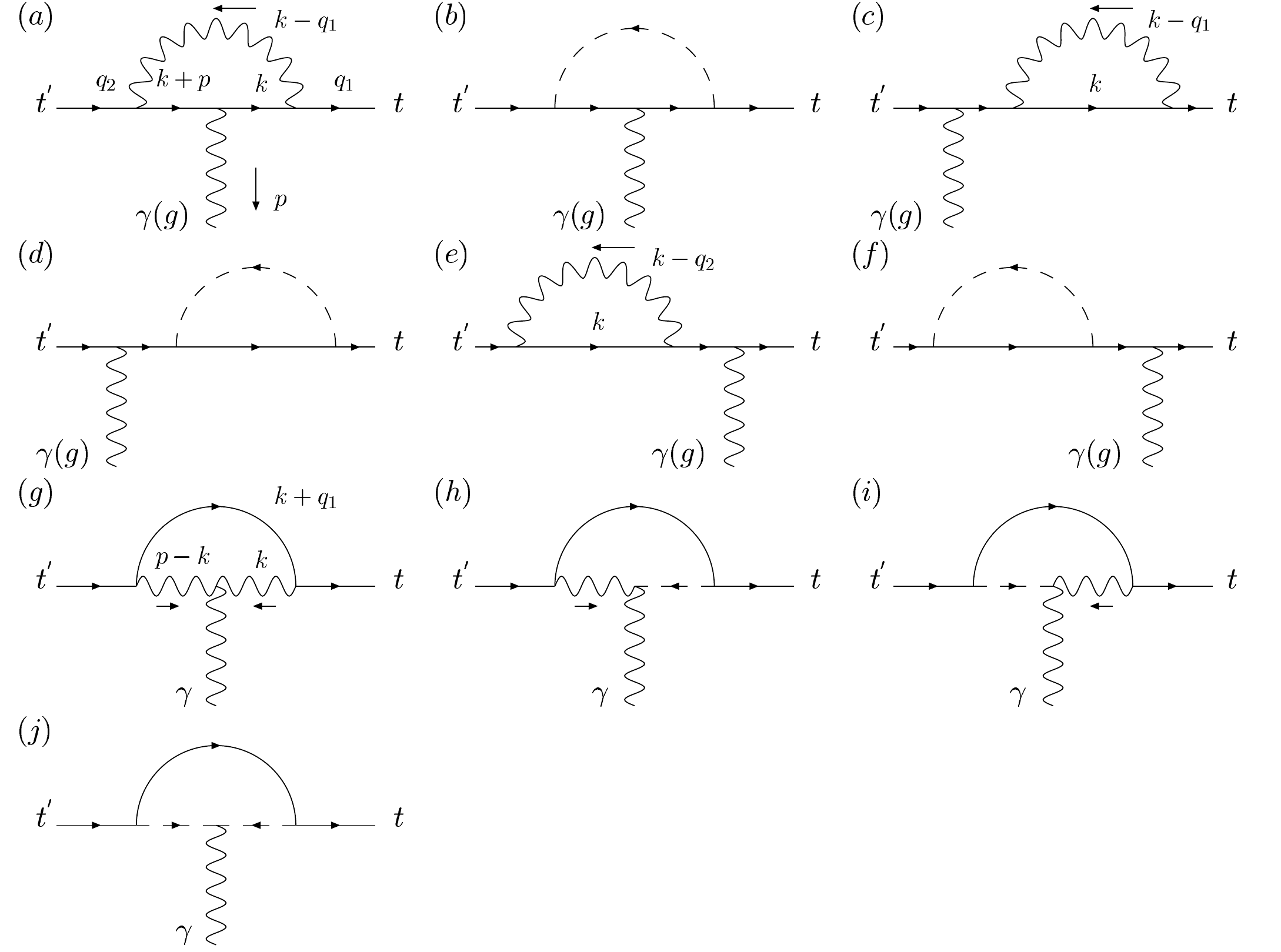,width=.90\textwidth}
\caption{Feynman diagrams contributing $t^{\prime} \to t\gamma (t g)$ process.
Solid lines represent quarks, internal (external) wavy lines correspond to gauge boson (photon), and dashed lines are scalars.}
\label{diagrams}
\end{center}
\end{figure}

We will assume that the vector $V$ in the loop, with mass $M_V$, has both left- and right-handed couplings with a generic pair of fermions:
\beq
\bar f\; \gamma^\mu V_\mu\, (g_{Lff'} L + g_{Rff'} R )\; f'\,;
\eeq
while a scalar boson $\phi$ has couplings:
\beq
\bar f\; (\lambda_{L ff'} L + \lambda_{Rff'} R)\, \phi\; f'\,.
\eeq
In our calculation, we use 't Hooft-Feynman gauge and take the external quarks $t^{\prime}$, $t$ to be on shell.
The $t^{\prime} \to t\gamma (t g)$ transition matrix element is given by
\begin{equation}
{\cal M} = \bar{t}(q_{1}) \Gamma_{{\rm ren}}^{\mu} t^{\prime}(q_{2}) \epsilon_{\mu}(\lambda),
\end{equation}
where $\Gamma_{{\rm ren}}^{\mu}$ is renormalised electromagnetic (strong) current and $\epsilon_{\mu}$ is the polarisation vector of the gauge boson.
In this case, the form factor of electromagnetic current is given by
\begin{equation}
\Gamma_{{\rm ren}}^{\mu} = \Gamma^{\mu} + T_{L}\gamma^{\mu}L + T_{R}\gamma^{\mu}R.
\end{equation}
The unrenormalised current $\Gamma^{\mu}$ can be written in the following form
\begin{eqnarray}
\Gamma^{\mu} = (a_{1}p^{\mu} + a_{2}q_{2}^{\mu} + a_{3}\gamma^{\mu})L + (b_{1}p^{\mu} + b_{2}q_{2}^{\mu} + b_{3}\gamma^{\mu})R,
\end{eqnarray}
where the form factors $a_{i}$ and $b_{i}$ are sums of the contributions from all diagrams in the process.
We perform our loop calculations in $D$ dimensions and define $\epsilon = 4 -D$.
In general, $\Gamma^{\mu}$ contains $1/\epsilon$ terms and has to be renormalised.
Detail of the renormalisation scheme is given in Ref.~\cite{Deshpande:1981zq}.
We introduce $T_{L}$ and $T_{R}$ which are left- and right-handed parts of the counter terms.
In the case of $t^{\prime} \to t\gamma$ where $\gamma$ is on shell (and similarly for the gluon), electromagnetic current conservation $p_{\mu} \Gamma_{{\rm ren}}^{\mu} =0$ implies the condition
\begin{eqnarray}
T_{L} &=& -a_{3} - \frac{1}{2}(m_{t}a_{2} + m_{t^{\prime}}b_{2}), \\
T_{R} &=& -b_{3} - \frac{1}{2}(m_{t^{\prime}}a_{2} + m_{t}b_{2}).
\end{eqnarray}
In the rest of this appendix we list the contribution of each diagram to the amplitude. This is a generalisation of the amplitudes in Ref.~\cite{Deshpande:1981zq}.
The currents $A^{\mu}$-$J^{\mu}$ denote the results for each diagrams labelled in Fig~\ref{diagrams}, expressed in terms of Passarino-Veltman functions $B_{0}$, $B_{1}$, $C_{0}$, $\tilde{C}_{0}$ and $C_{ij}$ which are given in Ref.~\cite{Passarino:1978jh}, so that the total result $\Gamma^\mu$ is the sum of all those cotributions.
In the following expressions, $M_{q}$ and $M_{V}$ are the masses of the internal quark $q$ and vector boson $V$ respectively, $Q_u = 2/3$ and $Q_d = -1/3$ the electric charges of up and down type quarks (note that in the gluon amplitude they are replaced by a colour factor $4/3$).
The values of the coefficients $A$, $B$, $C$ and $D$ are listed in Table~\ref{tab:coefficients}.
\begin{table}[h]
\begin{center}
\begin{tabular}{|cc||cccc|}
\hline
process & internal gauge (scalar) boson & $A$ & $B$ & $C$ & $D$ \\
\hline
$\gamma t$ & $W$ and $\phi_{W}$ &  $Q_{u}e$ & $Q_{q}e$ & $Q_{V}e$ & $gm_{W}\sin\theta_{W}$ \\
 & $Z$ and $\phi_{Z}$ & $Q_{u}e$ & $Q_{u}e$ & $0$ & $0$ \\
 & $H$ & $Q_{u}e$ & $Q_{u}e$ & $0$ & $0$ \\\hline
$gt$ & $W/Z/H$ & $g_{s}$ & $g_{s}$ & $0$ & $0$ \\
\hline
\end{tabular}
\end{center}
\caption{Values of the coefficients in the form factors $A^{\mu}$-$J^{\mu}$.
$\phi_{V}$ is the Goldstone boson. $Q_{V}$ ($Q_{q}$) is the electric charge of the internal gauge boson (quark).} \label{tab:coefficients}
\end{table}

The contribution of diagram $a$ is given by
\begin{eqnarray}
A^{\mu} &=& \frac{A}{16\pi^{2}}
\Bigl(
[
g_{Ltq}g_{Lqt^{\prime}}\, f_{A1}
+ g_{Rtq}g_{Rqt^{\prime}}\, f_{A4}
+ g_{Rtq}g_{Lqt^{\prime}}\, f_{A7}
]\; p^{\mu}L \nonumber \\
&& +[
g_{Ltq}g_{Lqt^{\prime}}\, f_{A2}
+ g_{Rtq}g_{Rqt^{\prime}}\, f_{A5}
+ g_{Rtq}g_{Lqt^{\prime}}\, f_{A8}
] \; q_{2}^{\mu}L \nonumber \\
&& +[
g_{Ltq}g_{Lqt^{\prime}}\, f_{A3}
+ g_{Rtq}g_{Rqt^{\prime}}\, f_{A6}
+ g_{Rtq}g_{Lqt^{\prime}}\, f_{A9}
+ g_{Ltq}g_{Rqt^{\prime}}\, f_{A10}
]\;  \gamma^{\mu}L
\Bigr) \nonumber \\
&& + (L \leftrightarrow R)\,,
\end{eqnarray}
where the coefficients $f$, which are functions of the masses, are:
%
%
\begin{equation} \begin{array}{rcl}
f_{A1}(M_{q},M_{V}) &=& -2m_{t}(\epsilon - 2)(C_{11} + C_{21})\,, \\
f_{A2}(M_{q},M_{V}) &=& 2 m_{t} [\epsilon C_{12} + (\epsilon - 2)C_{23}]\,, \\
f_{A3}(M_{q},M_{V}) &=& (-2m_{t^{\prime}} + \epsilon p^{2} - \epsilon m_{t}^{2})C_{12} + (\epsilon - 2)(2C_{24} - \tilde{C}_{0} - p^{2}C_{11})\nonumber \\
&& + (\epsilon - 2)M_{q}^{2}C_{0} \,,\\
f_{A4}(M_{q},M_{V}) &=& 2m_{t^{\prime}} [(\epsilon -2)(C_{11} - C_{23} + C_{21}) - \epsilon C_{12} ] \,,\\
f_{A5}(M_{q},M_{V}) &=& 2 m_{t^{\prime}} [2C_{12} + (\epsilon - 2)(C_{22} - C_{23})] \,,\\
f_{A6}(M_{q},M_{V}) &=& - m_{t}m_{t^{\prime}}(\epsilon + 2)C_{12} \,,\\
f_{A7}(M_{q},M_{V}) &=& M_{q}[2(4 - \epsilon)C_{11} + 2(2 - \epsilon)C_{0} ] \,,\\
f_{A8}(M_{q},M_{V}) &=& M_{q}[2(\epsilon -4)C_{12} + 2\epsilon C_{0} ] \,,\\
f_{A9}(M_{q},M_{V}) &=& -\epsilon M_{q} m_{t} C_{0} \,,\\
f_{A10}(M_{q},M_{V}) &=& -\epsilon M_{q} m_{t^{\prime}}C_{0}\,,
\end{array} \end{equation}
with $C_x$ being Passarino-Veltman functions evaluated in $C_x = C_{x}(p,-q_{2},M_{q},M_{q},M_{V})$; diagram $b$ contributes
\begin{eqnarray}
B^{\mu} &=& \frac{A}{16\pi^{2}}
\Bigl(
[
\lambda_{Ltq}\lambda_{Rqt^{\prime}}\, f_{B1}
+ \lambda_{Rtq}\lambda_{Lqt^{\prime}}\, f_{B4}
+ \lambda_{Ltq}\lambda_{Lqt^{\prime}}\, f_{B7}
]\; p^{\mu}L \nonumber \\
&& +[
\lambda_{Ltq}\lambda_{Rqt^{\prime}}\, f_{B2}
+ \lambda_{Rtq}\lambda_{Lqt^{\prime}}\, f_{B5}
+ \lambda_{Ltq}\lambda_{Lqt^{\prime}}\, f_{B8}
]\; q_{2}^{\mu}L \nonumber \\
&& +[
\lambda_{Ltq}\lambda_{Rqt^{\prime}}\, f_{B3}
+ \lambda_{Rtq}\lambda_{Lqt^{\prime}}\, f_{B6}
+ \lambda_{Ltq}\lambda_{Lqt^{\prime}}\, f_{B9}
+ \lambda_{Rtq}\lambda_{Rqt^{\prime}}\, f_{B10}
]\;  \gamma^{\mu}L
\Bigr) \nonumber \\
&& + (L \leftrightarrow R)\,,
\end{eqnarray}
with coefficients:
%
%
\begin{equation} \begin{array}{rcl}
f_{B1}(M_{q},M_{V}) &=& 2m_{t^{\prime}} (C_{12} + C_{23} - C_{11} - C_{21}) \,,\\
f_{B2}(M_{q},M_{V}) &=& 2m_{t^{\prime}} (C_{23} - C_{22}) \,,\\
f_{B3}(M_{q},M_{V}) &=& m_{t}m_{t^{\prime}} C_{12}\,, \\
f_{B4}(M_{q},M_{V}) &=& 2m_{t} (C_{11} + C_{21})\,, \\
f_{B5}(M_{q},M_{V}) &=& - 2m_{t} (C_{12} + C_{23}) \,,\\
f_{B6}(M_{q},M_{V}) &=& -M_{q}^{2}C_{0} + p^{2}C_{11} + (m_{t}^{2} - p^{2})C_{12} + \tilde{C}_{0} -2C_{24}\,, \\
f_{B7}(M_{q},M_{V}) &=& - 2M_{q}(C_{0} + C_{11})\,, \\
f_{B8}(M_{q},M_{V}) &=& 2M_{q}(C_{0} + C_{12})\,, \\
f_{B9}(M_{q},M_{V}) &=& - M_{q}m_{t}C_{0}\,, \\
f_{B10}(M_{q},M_{V}) &=& - M_{q}m_{t^{\prime}}C_{0}\,,
\end{array} \end{equation}
with $C_x = C_{x}(p,-q_{2},M_{q},M_{q},M_{V})$; diagram $c$ gives
\begin{eqnarray}
C^{\mu} &=& \frac{B}{16\pi^{2}}
[
g_{Ltq}g_{Lqt^{\prime}}\, f_{C1} 
+ g_{Rtq}g_{Rqt^{\prime}} \, f_{C2} + 
g_{Rtq}g_{Lqt^{\prime}}\,  f_{C3} + g_{Ltq}g_{Rqt^{\prime}}\, f_{C4}
]\; \gamma^{\mu}L \nonumber \\
&& + (L \leftrightarrow R)\,,
\end{eqnarray}
where the coefficients are equal to:
\begin{equation} \begin{array}{rcl}
f_{C1}(M_{q},M_{V}) &=& \frac{(\epsilon -2) m_{t}^{2} B_{1}}{(m_{t^{\prime}}^{2} - m_{t}^{2})}\,, \\
f_{C2}(M_{q},M_{V}) &=& \frac{(\epsilon -2) m_{t} m_{t^{\prime}} B_{1}}{(m_{t^{\prime}}^{2} - m_{t}^{2})} \,,\\
f_{C3}(M_{q},M_{V}) &=& \frac{(\epsilon -4) M_{q} m_{t} B_{0}}{(m_{t^{\prime}}^{2} - m_{t}^{2})} \,,\\
f_{C4}(M_{q},M_{V}) &=& \frac{(\epsilon -4) M_{q} m_{t^{\prime}} B_{0}}{(m_{t^{\prime}}^{2} - m_{t}^{2})}\,,
\end{array} \end{equation}
where $B_i$ are Passarino-Veltman functions evaluated at $B_i = B_{i}(-q_{1},M_{q},M_{V})$; diagram $d$ contributes
\begin{eqnarray}
D^{\mu} &=& \frac{B}{16\pi^{2}}
[
\lambda_{Ltq}\lambda_{Rqt^{\prime}}\,  f_{D1} + 
\lambda_{Rtq}\lambda_{Lqt^{\prime}}\, f_{D2} + 
\lambda_{Ltq}\lambda_{Lqt^{\prime}}\, f_{D3} + 
\lambda_{Rtq}\lambda_{Rqt^{\prime}}\, f_{D4}
] \; \gamma^{\mu}L \nonumber \\
&& + (L \leftrightarrow R)\,,
\end{eqnarray}
where the coefficients equal to:
%
%
\begin{equation} \begin{array}{rcl}
f_{D1}(M_{q},M_{V}) &=& \frac{m_{t}m_{t^{\prime}}B_{1}}{m_{t}^{2} - m_{t^{\prime}}^{2}} \,,\\
f_{D2}(M_{q},M_{V}) &=& \frac{m_{t}^{2}B_{1}}{m_{t}^{2} - m_{t^{\prime}}^{2}}\,, \\
f_{D3}(M_{q},M_{V}) &=& - \frac{M_{q}m_{t}B_{0}}{m_{t}^{2} - m_{t^{\prime}}^{2}} \,,\\
f_{D4}(M_{q},M_{V}) &=& - \frac{M_{q}m_{t^{\prime}}B_{0}}{m_{t}^{2} - m_{t^{\prime}}^{2}}\,,
\end{array} \end{equation}
with $B_i = B_{i}(-q_{1},M_{q},M_{V})$; the contribution of diagram $e$ can be parametrised as
\begin{eqnarray}
E^{\mu} &=& \frac{B}{16\pi^{2}}
[
g_{Ltq}g_{Lqt^{\prime}}\, f_{E1} + 
g_{Rtq}g_{Rqt^{\prime}}\, f_{E2} + 
g_{Rtq}g_{Lqt^{\prime}}\, f_{E3} + 
g_{Ltq}g_{Rqt^{\prime}} f_{E4}
]\; \gamma^{\mu}L \nonumber \\
&& + (L \leftrightarrow R)\,,
\end{eqnarray}
where
\begin{equation} \begin{array}{rcl}
f_{E1}(M_{q},M_{V}) &=& \frac{(\epsilon -2) m_{t^{\prime}}^{2} B_{1}}{(m_{t}^{2} - m_{t^{\prime}}^{2})} \,,\\
f_{E2}(M_{q},M_{V}) &=& \frac{(\epsilon -2) m_{t} m_{t^{\prime}} B_{1}}{(m_{t}^{2} - m_{t^{\prime}}^{2})}\,, \\
f_{E3}(M_{q},M_{V}) &=& \frac{(\epsilon -4) M_{q} m_{t} B_{0}}{(m_{t}^{2} - m_{t^{\prime}}^{2})}\,, \\
f_{E4}(M_{q},M_{V}) &=& \frac{(\epsilon -4) M_{q} m_{t^{\prime}} B_{0}}{(m_{t}^{2} - m_{t^{\prime}}^{2})}\,,
\end{array} \end{equation}
with $B_i = B_{i}(-q_{2},M_{q},M_{V})$; diagram $f$ gives
\begin{eqnarray}
F^{\mu} &=& \frac{B}{16\pi^{2}}
[
\lambda_{Ltq}\lambda_{Rqt^{\prime}}\, f_{F1} + \lambda_{Rtq}\lambda_{Lqt^{\prime}}\, f_{F2} + \lambda_{Ltq}\lambda_{Lqt^{\prime}} \, f_{F3} + \lambda_{Rtq}\lambda_{Rqt^{\prime}}\, f_{F4}
]\; \gamma^{\mu}L \nonumber \\
&& + (L \leftrightarrow R)\,,
\end{eqnarray}
where the coefficients are
%
\begin{equation} \begin{array}{rcl}
f_{F1}(M_{q},M_{V}) &=& \frac{m_{t}m_{t^{\prime}}B_{1}}{m_{t^{\prime}}^{2} - m_{t}^{2}}\,, \\
f_{F2}(M_{q},M_{V}) &=& \frac{m_{t^{\prime}}^{2}B_{1}}{m_{t^{\prime}}^{2} - m_{t}^{2}} \,,\\
f_{F3}(M_{q},M_{V}) &=& - \frac{M_{q}m_{t}B_{0}}{m_{t^{\prime}}^{2} - m_{t}^{2}}\,, \\
f_{F4}(M_{q},M_{V}) &=& - \frac{M_{q}m_{t^{\prime}}B_{0}}{m_{t^{\prime}}^{2} - m_{t}^{2}}\,,
\end{array} \end{equation}
with $B_i = B_{i}(-q_{2},M_{q},M_{V})$; diagram $g$ evaluates to 
\begin{eqnarray}
G^{\mu} &=& \frac{C}{16\pi^{2}} \Bigl(
[
g_{Ltq}g_{Lqt^{\prime}}\, f_{G1}
+ g_{Rtq}g_{Rt^{\prime}}\, f_{G4}
+ g_{Rtq}g_{Lt^{\prime}}\, f_{G7}
]\; p^{\mu}L \nonumber \\
&& +[
g_{Ltq}g_{Lqt^{\prime}}\, f_{G2}
+ g_{Rtq}g_{Rqt^{\prime}}\,  f_{G5}
+ g_{Rtq}g_{Lqt^{\prime}}\, f_{G8}
]\; q_{2}^{\mu}L \nonumber \\
&& +[
g_{Ltq}g_{Lqt^{\prime}}\, f_{G3}
+ g_{Rtq}g_{Rqt^{\prime}}\, f_{G6}
+ g_{Rtq}g_{Lqt^{\prime}}\, f_{G9}
+ g_{Ltq}g_{Rqt^{\prime}}\, f_{G10}
] \; \gamma^{\mu}L
\Bigr) \nonumber \\
&& + (L \leftrightarrow R)\,,
\end{eqnarray}
with coefficients
%
%
\begin{equation} \begin{array}{rcl}
f_{G1}(M_{q},M_{V}) &=& (\epsilon -2)m_{t} (C_{0} + 3C_{11} + 2C_{21})\,, \\
f_{G2}(M_{q},M_{V}) &=& 2m_{t}[2C_{0} + C_{11} + C_{12} - (\epsilon -2)(C_{12} + C_{23})] \,,\\
f_{G3}(M_{q},M_{V}) &=& m_{t}^{2}(C_{11} - C_{0}) - m_{t^{\prime}}^{2}(2C_{0} + C_{11} + C_{12}) + 2p^{2}(C_{0} + C_{11}) + 2\tilde{C}_{0} \nonumber \\
&& -2(\epsilon - 2)C_{24}\,, \\
f_{G4}(M_{q},M_{V}) &=& m_{t^{\prime}} [2C_{11} - 2C_{0} - 4C_{12} - (\epsilon - 2)(C_{11} - C_{12} + 2C_{21} - 2C_{23})]\,, \\
f_{G5}(M_{q},M_{V}) &=& 2m_{t^{\prime}} [C_{0} - C_{11} + 2C_{12} + (\epsilon - 2)(C_{23} - C_{22})] \,,\\
f_{G6}(M_{q},M_{V}) &=& - 3 m_{t}m_{t^{\prime}} (C_{0} + C_{12})\,, \\
f_{G7}(M_{q},M_{V}) &=& M_{q} [2C_{0} - 2C_{11} + (4 - \epsilon)(C_{0} + 2C_{11})] \,,\\
f_{G8}(M_{q},M_{V}) &=& 2M_{q} [C_{12} - 3C_{0} - (4 - \epsilon)C_{12}] \,,\\
f_{G9}(M_{q},M_{V}) &=& 3M_{q}m_{t}C_{0} \,,\\
f_{G10}(M_{q},M_{V}) &=& 3M_{q}m_{t^{\prime}}C_{0} \,,\\
\end{array} \end{equation}
with Passarino-Veltman functions $C_x = C_{x}(-p,q_{2},M_{V},M_{V},M_{q})$; the contribution of diagram $h$ is
\begin{eqnarray}
H^{\mu} &=& \frac{D}{16\pi^{2}} \Bigl(
[
\lambda_{Ltq}g_{Lqt^{\prime}}\, f_{H1}
]\; q_{2}^{\mu}L \nonumber \\
&& + [
\lambda_{Ltq}g_{Lqt^{\prime}}\, f_{H2}
+ \lambda_{Rtq}g_{Rqt^{\prime}}\, f_{H3}
+ \lambda_{Rtq}g_{Lqt^{\prime}}\, f_{H4}
]\; \gamma^{\mu}L
\Bigr) + (L \leftrightarrow R)\,,
\end{eqnarray}
where the coefficients are given by
%
%
\begin{equation} \begin{array}{rcl}
f_{H1}(M_{q},M_{V}) &=& 2(C_{12} - C_{11})\,, \\
f_{H2}(M_{q},M_{V}) &=& m_{t}(C_{0} + C_{11})\,, \\
f_{H3}(M_{q},M_{V}) &=& m_{t^{\prime}} (C_{11} - C_{12})\,, \\
f_{H4}(M_{q},M_{V}) &=& M_{q}C_{0}\,,
\end{array} \end{equation}
with $C_x = C_{x}(-p,q_{2},M_{V},M_{V},M_{q})$; diagram $i$ gives
\begin{eqnarray}
I^{\mu} &=& \frac{D}{16\pi^{2}} \Bigl(
[
g_{Rtq}\lambda_{Lqt^{\prime}}\, f_{I2}
]\; (p^{\mu} - q_{2}^{\mu})L \nonumber \\
&& + [
g_{Ltq}\lambda_{Rqt^{\prime}}\, f_{I1}
+ g_{Rtq}\lambda_{Lqt^{\prime}}\, f_{I3}
+ g_{Ltq}\lambda_{Lqt^{\prime}}\, f_{I4}
]\; \gamma^{\mu}L
\Bigr) + (L \leftrightarrow R)\,,
\end{eqnarray}
where the coefficients are equal to
%
%
\begin{equation} \begin{array}{rcl}
f_{I1}(M_{q},M_{V}) &=& m_{t^{\prime}} (C_{12} - C_{11}) \,,\\
f_{I2}(M_{q},M_{V}) &=& - 2(C_{0} + C_{11}) \,,\\
f_{I3}(M_{q},M_{V}) &=& - m_{t} (C_{0} + C_{11}) \,,\\
f_{I4}(M_{q},M_{V}) &=& M_{q} C_{0}\,,
\end{array} \end{equation}
with $C_x = C_{x}(-p,q_{2},M_{V},M_{V},M_{q})$; finally, diagram $j$ contributes
\begin{eqnarray}
J^{\mu} &=& \frac{C}{16\pi^{2}} \Bigl(
[
\lambda_{Ltq}\lambda_{Rqt^{\prime}}\, f_{J1} + 
\lambda_{Rtq}\lambda_{Lqt^{\prime}}\, f_{J3} + 
\lambda_{Ltq}\lambda_{Lqt^{\prime}}\, f_{J6}
]\; p^{\mu}L \nonumber \\
&& + [
\lambda_{Ltq}\lambda_{Rqt^{\prime}}\, f_{J2} + 
\lambda_{Rtq}\lambda_{Lqt^{\prime}}\, f_{J4} + 
\lambda_{Ltq}\lambda_{Lqt^{\prime}}\, f_{J7}
]\; q_{2}^{\mu}L \nonumber \\
&& + [
\lambda_{Rtq}\lambda_{Lqt^{\prime}}\, f_{J5}
]\; \gamma^{\mu}L
\Bigr)
+ (L \leftrightarrow R)\,,
\end{eqnarray}
with coefficients
%
%
\begin{equation} \begin{array}{rcl}
f_{J1} &=& m_{t^{\prime}} (C_{11} - C_{12} + 2C_{21} - 2C_{23})\,, \\
f_{J2} &=& 2m_{t^{\prime}}(C_{22} - C_{23})\,, \\
f_{J3} &=& - m_{t} (C_{0} + 3C_{11} + 2C_{21})\,, \\
f_{J4} &=& 2m_{t} (C_{12} + C_{23})\,, \\
f_{J5} &=& 2C_{24} \,,\\
f_{J6} &=& - M_{q} (C_{0} + 2C_{11}) \,,\\
f_{J7} &=& 2M_{q} C_{12}\,,
\end{array} \end{equation}
with $C_x = C_{x}(-p,q_{2},M_{V},M_{V},M_{q})$.


\section{Decay width formulas}

\subsection{A quick estimate}
\label{sec:estim}

Just to have an idea of the orders of magnitude, we give the tree level widths by computing the decay of the heavy fermion in $W$ + light fermion (where the dominant channel is in longitudinal polarisation, i.e. in the Goldstone boson).
From Eq.~\ref{eq:treeGamma}, the width is:
\beq
\Gamma (F \to W f') \simeq \frac{\lambda^2 m_F}{32 \pi}\,,
\eeq
where $\lambda$ is the new Yukawa coupling.
This width is equal to the one into $Z$ (long polarisation) and neutral Higgs, therefore if both channels are present the total width is doubled.

For the loop induced one, the graphs directly proportional to the new Yukawa coupling dominate.
We will distinguish two cases: Yukawa coupling involving the left-handed $F$ (case I) and right-handed $F$ (case II).
The amplitudes are dominated by the terms:
\beq
\mbox{case I} & \Rightarrow & \left\{ \begin{array}{l}
m_F a_2 \simeq 0\\
m_F b_2 \simeq \frac{i \lambda \lambda_f e}{16 \pi^2} \sum \frac{\lambda_q}{\lambda_f} (Q_q f_{B1} + Q_V f_{J1} )
\end{array} \right. \\
\mbox{case II} & \Rightarrow & \left\{ \begin{array}{l}
m_F a_2 \simeq \frac{i \lambda \lambda_f e}{16 \pi^2} \sum (Q_q f_{B1} + Q_V f_{J1} )\\
m_F b_2 \simeq 0
\end{array} \right.
\eeq
Plugging those amplitudes in Eq.~\ref{eq:loopGamma}, we find:
\beq
\Gamma (F \to \gamma f) \simeq \frac{\lambda^2 m_F}{32 \pi}\, \frac{\alpha_{ew} \alpha_t}{32 \pi^2}\; | \sum |^2\,,
\eeq
where we have used the top Yukawa coupling in $\alpha_t = \lambda_t^2/4\pi$, and $\alpha_{em} = e^2/4\pi$. (Note: for light fermions $f$, the top Yukawa may be replaced by the weak $\alpha$, from diagrams H and I).
The branching fraction is therefore:
\beq
BR (F \to \gamma f) \sim \frac{\alpha_{ew} \alpha_t}{32 \pi^2}\; | \sum |^2 \simeq 10^{-6} \cdot | \sum |^2\,.
\eeq
Note that this is a very conservative estimate, as the sum may contain a $\log \frac{m_F^2}{m_W^2}$ enhancement which may enhance the result by one or two orders of magnitude.

\subsection{Decay widths and the the large mass limit}
\label{sec:widths}

The heavy vector-like fermion $t^{\prime}$ can decay at tree level to $b W$ and $b^{\prime} W$ via a charged current and to $Z t$ and $h t$ via a neutral current.
The $t^{\prime} \to bW$ and $t^{\prime} \to b^{\prime}W$ transition matrix elements are given by
\begin{equation}
{\cal M} = \bar{b}(q_{1}) \gamma^{\mu} (g_{L}^{W}L + g_{R}^{W}R) t^{\prime}(q_{2}) \epsilon_{\mu}(\lambda)\,,
\end{equation}
where $g_{L}^{W}$ and $g_{R}^{W}$ are left- and right-handed couplings of $t^{\prime} \bar{b}W$ ($t^{\prime}\bar{b}^{'}W$).
From Eq.~\ref{eq:treeGamma}, the partial width of $t^{\prime} \to bW (b^{\prime}W)$ decay is expressed as
\begin{eqnarray}
\Gamma (t^{\prime} \to bW) &=& \frac{\lambda^{\frac{1}{2}}(1,\frac{m_{b}^{2}}{m_{t^{\prime}}^{2}},\frac{m_{W}^{2}}{m_{t^{\prime}}^{2}})}{32\pi m_{t^{\prime}}} \Biggl\{ (|g_{L}^{W}|^{2} + |g_{R}^{W}|^{2}) \left[ m_{t^{\prime}}^{2} + m_{b}^{2} - 2m_{W}^{2} + \frac{(m_{t^{\prime}}^{2} - m_{b}^{2})^{2}}{m_{W}^{2}} \right] \nonumber \\
&& - 12 (Re\, g_{L}^{W} Re\, g_{R}^{W} + Im\, g_{L}^{W} Im\, g_{R}^{W})\,  m_{t^{\prime}}m_{b} \Biggr\}
\end{eqnarray}
where $Re$ and $Im$ indicate respectively the real and imaginary part.
In the large $m_{t^{\prime}}$ limit the previous formula becomes
\begin{equation}
\Gamma (t^{\prime} \to b^{(\prime)} W) \sim \frac{|g_{L}^{W}|^{2} + |g_{R}^{W}|^{2}}{32\pi}  \frac{m_{t^\prime}^3}{m_{W}^{2}} \left(1  - \frac{m_{b^{(\prime)}}^{2}}{m_{t^\prime}^2}  \right)^2.
\end{equation}

Concerning neutral currents, the $t^{\prime} \to tZ$ transition matrix element is written as
\begin{equation}
{\cal M} = \bar{t}(q_{1}) \gamma^{\mu} (g_{L}^{Z}L + g_{R}^{Z}R) t^{\prime}(q_{2}) \epsilon_{\mu}(\lambda)
\end{equation}
where $g_{L}^{Z}$ and $g_{R}^{Z}$ are left- and right-handed couplings of $t^{\prime} \bar{t} Z$.
The partial width of $t^{\prime} \to tZ$ is
\begin{eqnarray}
\Gamma (t^{\prime} \to tZ) &=& \frac{\lambda^{\frac{1}{2}}(1,\frac{m_{t}^{2}}{m_{t^{\prime}}^{2}},\frac{m_{Z}^{2}}{m_{t^{\prime}}^{2}})}{32\pi m_{t^{\prime}}} \Biggl\{ (|g_{L}^{Z}|^{2} + |g_{R}^{Z}|^{2}) \left[ m_{t^{\prime}}^{2} + m_{t}^{2} - 2m_{Z}^{2} + \frac{(m_{t^{\prime}}^{2} - m_{t}^{2})^{2}}{m_{Z}^{2}} \right] \nonumber \\
&& - 12 (Re\, g_{L}^{Z} Re\, g_{R}^{Z} + Im\, g_{L}^{Z} Im\, g_{R}^{Z}) m_{t^{\prime}}m_{t} \Biggr\}\; .
\end{eqnarray}
In the large $t'$ mass limit the previous formula can be written in the following form
\begin{equation}
\Gamma (t^{\prime} \to t Z) \sim \frac{|g_{L}^{Z}|^{2} + |g_{R}^{Z}|^{2}}{32\pi}  \frac{m_{t^\prime}^3}{m_{Z}^{2}}.
\end{equation}

For the tree level $t^{\prime} \to t h$ decay, the matrix element is written as
\begin{equation}
{\cal M} = \bar{t}(q_{1}) (C_{L}L + C_{R}R) t^{\prime}(q_{2})
\end{equation}
where $C_{L}$ and $C_{R}$ are left- and right-handed couplings of $t^{\prime} \bar{t} h$.
The partial width of $t^{\prime} \to t h$
\begin{eqnarray}
\Gamma (t^{\prime} \to t h) &=& \frac{\lambda^{\frac{1}{2}}(1,\frac{m_{t}^{2}}{m_{t^{\prime}}^{2}},\frac{m_{H}^{2}}{m_{t^{\prime}}^{2}})}{32\pi m_{t^{\prime}}} \biggl\{ (m_{t^{\prime}}^{2} + m_{t}^{2} - m_{H}^{2}) ( |C_{L}|^{2} + |C_{R}|^{2} ) \nonumber \\
&& + 4 m_{t^{\prime}}m_{t} (Re\, C_{L} Re\, C_{R} + Im\, C_{L} Im\, C_{R}) \biggr\}\,.
\end{eqnarray}
In the large $m_{t^{\prime}}$ limit this becomes
\begin{equation}
\Gamma (t^{\prime} \to t h) \sim \frac{ |C_{L}|^{2} + |C_{R}|^{2} }{32\pi}  m_{t^\prime}.
\end{equation}

\subsubsection*{Singlet case}

In the case the new vector-like fermion is a singlet, from the results in section~\ref{sec:mix2} we find that the couplings can be written as
\begin{eqnarray}
g_{L}^{W} &=& \frac{g}{\sqrt{2}} \sin \theta_{u}^{L} \sim \frac{g}{\sqrt{2}} \frac{x}{M}\,, \nonumber\\
g_{R}^{W} &=& 0\,, \nonumber\\
g_{L}^{Z} &=& \frac{g}{2\cos\theta_{W}} \cos \theta_{u}^{L} \sin \theta_{u}^{L}  \sim \frac{g}{2\cos \theta_{W}} \frac{x}{M}\,,  \\
g_{R}^{Z} &=& 0\,, \nonumber\\
C_{L} &=& - \frac{g}{2m_{W}} m_t \cos \theta_u^L \sin \theta_u^L \sim - \frac{g}{2m_{W}} \frac{m_{t}x}{M}\,,  \nonumber\\
C_{R} &=& - \frac{g}{2m_{W}}  m_{t'} \cos \theta_u^L \sin \theta_u^L \sim - \frac{g}{2m_{W}} x\,. \nonumber
\end{eqnarray}
In the large mass limit $m_{t'} \sim M \gg m_t, x$, the tree level $t^{\prime}$ decay widths become
\begin{eqnarray}
\Gamma (t^{\prime} \to bW) &\sim& \frac{1}{32\pi M} \frac{g^{2}}{2m_{W}^{2}} x^{2}M^{2}\,, \\
\Gamma (t^{\prime} \to tZ) &\sim& \frac{1}{32\pi M} \frac{g^{2}}{4m_{Z}^{2}\cos^{2}\theta_{W}} x^{2}M^{2}\,, \\
\Gamma (t^{\prime} \to th) &\sim& \frac{1}{32\pi M} \frac{g^{2}}{4m_{W}^{2}} x^{2}M^{2}\,;
\end{eqnarray}
with ratios
\begin{eqnarray}
\Gamma (t^{\prime} \to bW) / \Gamma (t^{\prime} \to tZ) &=& \frac{2m_{Z}^{2}\cos^{2}\theta_{W}}{m_{W}^{2}} = 2\,,\\
\Gamma (t^{\prime} \to th) / \Gamma (t^{\prime} \to tZ) &=& \frac{m_{Z}^{2}\cos^{2}\theta_{W}}{m_{W}^{2}} = 1\,.
\end{eqnarray}
Therefore, in this limit, the $t'$ will decay half of the times in $W b$, and 25\% in $Z t$ and 25\% in $h t$.

\subsubsection*{SM doublet case}

For a Standard Model like fermion doublet the couplings are given by:
\begin{eqnarray}
g_{L}^{W} &=& \frac{g}{\sqrt{2}} \sin (\theta_{u}^{L} - \theta_{d}^{L}) \sim \frac{g}{\sqrt{2}} \left(\frac{m_{t}x}{M^{2}} - \frac{m_{b}x_{b}}{M^{2}}\right)\,, \nonumber \\
g_{R}^{W} &=& -\frac{g}{\sqrt{2}} \cos\theta_{u}^{R} \sin\theta_{d}^{R} \sim -\frac{g}{\sqrt{2}} \frac{x_{b}}{M} \,,\nonumber \\
g_{L}^{Z} &=& 0\,, \\
g_{R}^{Z} &=& -\frac{g}{2\cos\theta_{W}} \cos \theta_{u}^{R} \sin \theta_{u}^{R}  \sim -\frac{g}{2\cos \theta_{W}} \frac{x}{M} \,,\nonumber \\
C_{L} &=& - \frac{g}{2m_{W}}  m_{t'} \cos \theta_u^R \sin \theta_u^R \sim - \frac{g}{2m_{W}} x \,, \nonumber\\
C_{R} &=& - \frac{g}{2m_{W}} m_t \cos \theta_u^R \sin \theta_u^R \sim - \frac{g}{2m_{W}} \frac{m_{t}x}{M}\,; \nonumber
\end{eqnarray}
in the large mass limit, the tree level $t^{\prime}$ decay widths are
\begin{eqnarray}
\Gamma (t^{\prime} \to bW) &\sim& \frac{1}{32\pi M} \frac{g^{2}}{2m_{W}^{2}} x_{b}^{2}M^{2}\,, \\
\Gamma (t^{\prime} \to tZ) &\sim& \frac{1}{32\pi M} \frac{g^{2}}{4m_{Z}^{2}\cos^{2}\theta_{W}} x^{2}M^{2}\,, \\
\Gamma (t^{\prime} \to th) &\sim& \frac{1}{32\pi M} \frac{g^{2}}{4m_{W}^{2}} x^{2}M^{2}\,;
\end{eqnarray}
with ratios
\begin{eqnarray}
\Gamma (t^{\prime} \to bW) / \Gamma (t^{\prime} \to tZ) &=& \frac{2m_{Z}^{2}\cos^{2}\theta_{W}}{m_{W}^{2}} \left( \frac{x_{b}}{x} \right)^{2} = 2 \left( \frac{x_{b}}{x} \right)^{2}\,, \\
\Gamma (t^{\prime} \to th) / \Gamma (t^{\prime} \to tZ) &=& \frac{m_{Z}^{2}\cos^{2}\theta_{W}}{m_{W}^{2}} = 1\,.
\end{eqnarray}
For small $x_b \gg x$, the $t'$ will always decay in tops and in equal proportions to $Z$ and Higgs.

\subsubsection*{Non SM doublet case}

For a doublet with hypercharge $7/6$, the couplings are
\begin{eqnarray}
g_{L}^{W} &=& \frac{g}{\sqrt{2}} \sin \theta_{u}^{L} \sim \frac{g}{\sqrt{2}} \frac{m_{t}x}{M^{2}}\,,  \nonumber\\
g_{R}^{W} &=& 0\,,  \nonumber\\
g_{L}^{Z} &=& \frac{g}{\cos\theta_{W}} \sin \theta_{u}^{L} \cos \theta_{u}^{L}  \sim \frac{g}{\cos \theta_{W}} \frac{xm_{t}}{M^{2}}\,, \\
g_{R}^{Z} &=& \frac{g}{2\cos\theta_{W}} \sin \theta_{u}^{R} \cos \theta_{u}^{R}  \sim \frac{g}{2\cos \theta_{W}} \frac{x}{M}\,, \nonumber \\
C_{L} &=& - \frac{g}{2m_{W}}  m_{t'} \cos \theta_u^R \sin \theta_u^R \sim - \frac{g}{2m_{W}} x \,, \nonumber\\
C_{R} &=& - \frac{g}{2m_{W}} m_t \cos \theta_u^R \sin \theta_u^R \sim - \frac{g}{2m_{W}} \frac{m_{t}x}{M}\,; \nonumber
\end{eqnarray}
In the large $M$ limit, the tree level $t^{\prime}$ decay widths are
\begin{eqnarray}
\Gamma (t^{\prime} \to bW) &\sim& \frac{1}{32\pi M} \frac{g^{2}}{2m_{W}^{2}} m_{t}^{2} x^{2}\,, \\
\Gamma (t^{\prime} \to tZ) &\sim& \frac{1}{32\pi M} \frac{g^{2}}{4m_{Z}^{2}\cos^{2}\theta_{W}} x^{2}M^{2}\,, \\
\Gamma (t^{\prime} \to th) &\sim& \frac{1}{32\pi M} \frac{g^{2}}{4m_{W}^{2}} x^{2}M^{2}\,;
\end{eqnarray}
with ratios
\begin{eqnarray}
\Gamma (t^{\prime} \to bW) / \Gamma (t^{\prime} \to tZ) &=& \frac{2m_{Z}^{2}\cos^{2}\theta_{W}}{m_{W}^{2}} \frac{m_{t}^{2}}{M^{2}}  = 2 \frac{m_{t}^{2}}{M^{2}} \sim 0\,, \\
\Gamma (t^{\prime} \to th) / \Gamma (t^{\prime} \to tZ) &=& \frac{m_{Z}^{2}\cos^{2}\theta_{W}}{m_{W}^{2}} = 1\,.
\end{eqnarray}
The $t'$ will therefore decay preferably in tops, with equal probability in association of a $Z$ boson or Higgs.

\subsubsection*{Triplet $Y=2/3$ case}

For a triplet with hypercharge $2/3$:
\begin{eqnarray}
g_{L}^{W} &=& \frac{g}{\sqrt{2}} \left( \cos \theta_{d}^{L} \sin \theta_{u}^{L} - \sqrt{2} \sin \theta_{d}^{L} \cos \theta_{u}^{L} \right)
\sim - \frac{g}{\sqrt{2}} \frac{x}{M} \,,  \nonumber\\
g_{R}^{W} &=& \frac{g}{\sqrt{2}} \left(- \sqrt{2} \sin \theta_{d}^{R} \cos \theta_{u}^{R} \right)
\sim - \frac{g}{\sqrt{2}} \frac{2xm_{b}}{M^{2}}\,,  \nonumber\\
g_{L}^{Z} &=& \frac{g}{2\cos\theta_{W}} \sin \theta_{u}^{L} \cos \theta_{u}^{L}  \sim \frac{g}{2\cos \theta_{W}} \frac{x}{M}\,, \\
g_{R}^{Z} &=& 0 \,, \nonumber\\
C_{L} &=& - \frac{g}{2m_{W}} m_t \sin \theta_{u}^{L} \cos \theta_{u}^{L} \sim - \frac{g}{2m_{W}} \frac{m_{t}x}{M} \,, \nonumber\\
C_{R} &=& - \frac{g}{2m_{W}} m_{t'} \sin \theta_{u}^{L} \cos \theta_{u}^{L} \sim- \frac{g}{2m_{W}} x\,. \nonumber
\end{eqnarray}
In the large $M$ limit, the tree level $t^{\prime}$ decay widths are
\begin{eqnarray}
\Gamma (t^{\prime} \to bW) &\sim& \frac{1}{32\pi M} \frac{g^{2}}{2m_{W}^{2}} x^{2}M^{2}\,, \\
\Gamma (t^{\prime} \to tZ) &\sim& \frac{1}{32\pi M} \frac{g^{2}}{4m_{Z}^{2}\cos^{2}\theta_{W}} x^{2}M^{2}\,, \\
\Gamma (t^{\prime} \to th) &\sim& \frac{1}{32\pi M} \frac{g^{2}}{4m_{W}^{2}} x^{2}M^{2}\,;
\end{eqnarray}
with ratios
\begin{eqnarray}
\Gamma (t^{\prime} \to bW) / \Gamma (t^{\prime} \to tZ) &=& \frac{2m_{Z}^{2}\cos^{2}\theta_{W}}{m_{W}^{2}} = 2\,, \\
\Gamma (t^{\prime} \to th) / \Gamma (t^{\prime} \to tZ) &=& \frac{m_{Z}^{2}\cos^{2}\theta_{W}}{m_{W}^{2}} = 1\,.
\end{eqnarray}


\subsubsection*{Triplet $Y=-1/3$ case}

In the case of a triplet with hypercharge $-1/3$:
\begin{eqnarray}
g_{L}^{W} &=& \frac{g}{\sqrt{2}} \left( \cos \theta_{d}^{L} \sin \theta_{u}^{L} + \sqrt{2} \sin \theta_{d}^{L} \cos \theta_{u}^{L} \right)
\sim \frac{g}{\sqrt{2}} \frac{\sqrt{2}x^{3}}{2M^{3}}\,, \nonumber\\
g_{R}^{W} &=& \frac{g}{\sqrt{2}} \left( \sqrt{2} \sin \theta_{d}^{R} \cos \theta_{u}^{R} \right)
\sim - \frac{g}{\sqrt{2}} \frac{\sqrt{2}xm_{b}}{M^{2}}\,, \nonumber \\
g_{L}^{Z} &=& -\frac{g}{2\cos\theta_{W}} \sin \theta_{u}^{L} \cos \theta_{u}^{L}  \sim -\frac{g}{2\cos \theta_{W}} \frac{\sqrt{2}x}{M} \,,\\
g_{R}^{Z} &=& -\frac{g}{\cos\theta_{W}} \cos \theta_{u}^{R} \sin \theta_{u}^{R} \sim -\frac{g}{\cos \theta_{W}} \frac{\sqrt{2}xm_{t}}{M^{2}}\,, \nonumber\\
C_{L} &=& - \frac{g}{2m_{W}} m_t \sin \theta_{u}^{L} \cos \theta_{u}^{L} \sim - \frac{g}{2m_{W}} \frac{\sqrt{2}xm_{t}}{M}\,, \\
C_{R} &=& - \frac{g}{2m_{W}}  m_{t'} \sin \theta_{u}^{L} \cos \theta_{u}^{L} \sim - \frac{g}{2m_{W}} \sqrt{2}x\,. \nonumber
\end{eqnarray}
In the large $M$ limit, the tree level $t^{\prime}$ decay widths are
\begin{eqnarray}
\Gamma (t^{\prime} \to bW) &\sim& \frac{1}{32\pi M} \frac{g^{2}}{2m_{W}^{2}} 2m_{b}^{2} x^{2}\,, \\
\Gamma (t^{\prime} \to tZ) &\sim& \frac{1}{32\pi M} \frac{g^{2}}{2m_{Z}^{2}\cos^{2}\theta_{W}} x^{2}M^{2}\,, \\
\Gamma (t^{\prime} \to th) &\sim& \frac{1}{32\pi M} \frac{g^{2}}{4m_{W}^{2}} 2x^{2}M^{2}\,;
\end{eqnarray}
with ratios
\begin{eqnarray}
\Gamma (t^{\prime} \to bW) / \Gamma (t^{\prime} \to tZ) &=& \frac{2m_{Z}^{2}\cos^{2}\theta_{W}}{m_{W}^{2}} \frac{m_{b}^{2}}{M^{2}}  = 2 \frac{m_{b}^{2}}{M^{2}} \sim 0\,, \\
\Gamma (t^{\prime} \to th) / \Gamma (t^{\prime} \to tZ) &=& \frac{m_{Z}^{2}\cos^{2}\theta_{W}}{m_{W}^{2}} = 1\,.
\end{eqnarray}

\end{document}